\documentclass[aps,prl,amsmath,twocolumn,groupedaddress,letterpaper,floatfix,longbibliography]{revtex4-2}

\usepackage{graphicx,subfigure,epsfig}
\usepackage{array,multirow}
\usepackage{dcolumn}
\usepackage{amssymb,amsmath,amsfonts,mathrsfs}
\usepackage{array}
\usepackage{newtxtext,setspace}
\usepackage{latexsym}
\usepackage{flafter,bm,bbm}
\usepackage{epstopdf,color,multirow}
\usepackage[colorlinks,linkcolor=blue,anchorcolor=blue,urlcolor=blue,citecolor=blue]{hyperref}
\usepackage{footnote}
\usepackage{booktabs}
\usepackage{soul}
\usepackage{cancel}

\makeatletter
\usepackage{dcolumn}
\usepackage{float}
\usepackage{bm}
\usepackage{dsfont}
\usepackage{multirow}
\usepackage{color}
\newcommand{\bra}[1]{\langle#1}
\newcommand{\ket}[1]{#1\rangle}
\newcommand{\vbra}[1]{\langle#1\rvert}
\newcommand{\vket}[1]{\lvert#1\rangle}
\newcommand{\average}[1]{\langle#1\rangle}
\newcommand{\blue}[1]{{\color{blue}{#1}}}

\begin{document}

\title{
    Disorder-induced Diffusion Transport in Flat-band Systems with Quantum Metric
}

\author{Chun Wang Chau$^{1,3}$}\thanks{These authors contributed equally to this work}
\author{Tian Xiang$^{1}$}\thanks{These authors contributed equally to this work}
\author{Shuai A. Chen$^{1,2}$}
\email{chsh@ust.hk}
\author{K. T. Law$^{1}$}
\email{phlaw@ust.hk}

\affiliation{1. Department of Physics, Hong Kong University of Science and Technology,
Clear Water Bay, Hong Kong, China}
\affiliation{2. Max Planck Institute for the Physics of Complex Systems, N\"{o}thnitzer Stra{\ss}e 38, Dresden 01187, Germany}
\affiliation{3. Cavendish Laboratory, Department of Physics, J J Thomson Avenue, Cambridge CB3 0HE, United Kingdom}
\date{\today}

\begin{abstract}
Our previous understanding of transport in disordered system depends on the assumption that there is a well-defined Fermi velocity. The Fermi velocity determines important length scales in the system such as the diffusion length and localization length. However, nearly flat band materials with vanishing Fermi velocity, it is uncertain how to understand the disorder effects and what quantities determine the characteristic length scales in the system. In the clean limit, it is expected that the bulk transport is absent. In this work, we demonstrate, with a diamond lattice, that disorder can induce diffusion transport in a flat-band system with finite quantum metric.  As disorder increases, the bulk transmission channels are activated, and the conductance reaches a maximum before decays inversely with disorder strength. Importantly, via the calculation of the wave-packet dynamics numerically, we show that the quantum metric determines the diffusion length of the system. Analytically, we show that the interplay between the disorder and quantum geometry gives rise to an effective Fermi velocity, as captured by the self-consistent Born approximation. The diffusion coefficient is identified from the Bethe-Salpeter equation under the ladder approximation. Our results reveal a disorder-driven delocalization mechanism in flat-band systems with finite quantum metric which cannot be understood by well-established theories of quantum diffusion. Our theory is important for understanding the disorder effects and transport properties of flat band materials such as twisted bilayer graphene which are current under intense investigation.  \end{abstract}

\maketitle

\begin{figure}[t]
\centering 
\includegraphics[width=1\columnwidth]{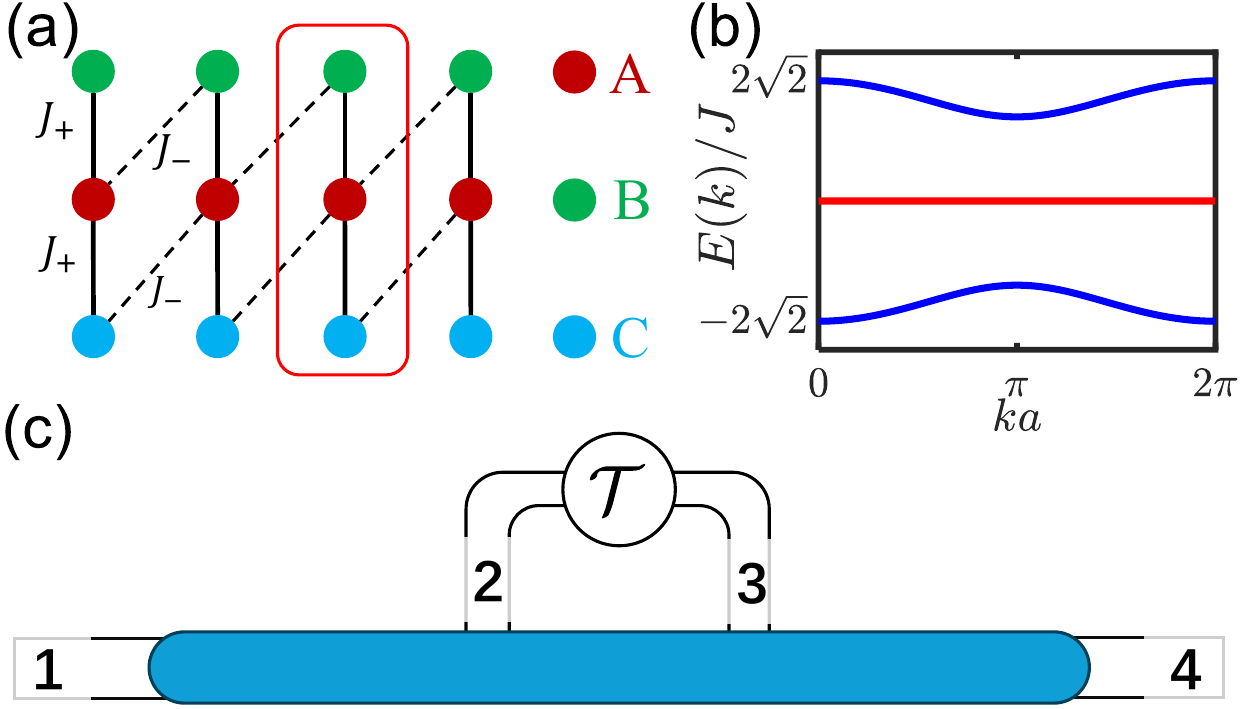}
\caption{
    (a) Schematic of the 1D diamond lattice, which contains three sites A, B, C per unit cell. (b)  Energy spectrum of the 1D diamond lattice. The central gap is exaggerated for clarity. (c) Schematic of the four-terminal M/FB/M junction. The central disordered 1D diamond chain (blue) of length $L$ serves as the device under measurement, with four metallic leads attached. Lead 1 and 4 are connected to the two ends of the chain, while lead 2 and 3 divide the chain into three segments, forming a $\pi$-shaped configuration. The total length of the disordered part is $L = L_{12} + L_{23} + L_{34}$, with $L_{12} = L_{34} = 10$ fixed throughout this work. Subscripts denote the corresponding lead labels as shown in (c).
    }
\label{Fig:junction} 
\end{figure}


\emph{\color{blue}Introduction.—} Flat-band systems, characterized by dispersionless energy bands, have recently gained significant attention. These materials emerged as a fertile platform for exploring diverse quantum phenomena including correlated insulating phases~\cite{CIP_1,CIP_2}, superconductivity~\cite{SC_1,SC_2,QM_FB_Josephson,FB-SC_1,FB-SC_2,FB-SC_3,FB-SC_4,FB-SC_5,FB_1,FB_2}, antiferromagnetism~\cite{FM_1,FM_2}, and excitonic effects~\cite{Ex_1,Ex_2}. The quantum geometric tensor, which quantifies the phase and amplitude distances between quantum states~\cite{QMT,QM_D_1,QM_D_2}, has emerged as a key quantity governing the physical properties of flat-band systems~\cite{QM_SC_1,QM_SC_2,FB_2}. For flat bands with nontrivial quantum geometric tensor, the zero-temperature conductivity is predicted to be related to the real part of the quantum geometric tensor—the quantum metric\cite{Kubo-Greenwood-1,QM-FB-Dis_1,Dis_1,FB-deg-1}.

In conventional band theory, partially filled dispersive bands yield metallic behavior with finite conductivity $\sigma$, as described by the Einstein relation $\sigma = e^2 D \rho(E)$, where $D$ is the diffusion coefficient and $\rho(E)$ is the density of states at the Fermi energy~\cite{MCMP-Girvin}. For Fermi liquids, $D = v_F^2 \tau$, with $v_F$ the Fermi velocity and $\tau$ the scattering time. While the density of states (DOS) sets the number of available carriers, $D$ characterizes their mobility and is linked to the Fermi velocity $v_F$. In contrast, non-interacting flat-band systems feature vanishing $v_F$ and diverging effective mass, leading to localized states and insulating behavior~\cite{Kubo-Greenwood-1}, which is consistent with the semiclassical picture where the vanishing group velocity precludes transport. This picture changes when additional mechanisms—such as inelastic scattering~\cite{Dis_1,Dis_2,FB-Nonzero-trans-1}, defects~\cite{QM-FB-Dis_1,QM-FB-Dis_2} and interactions~\cite{Int_1,Int_2,Int_3,FB-Localization-1}—are introduced. Recent experiments have demonstrated that disorder can induce delocalization for flat bands~\cite{FB-SC-Qbit-Array}. Nevertheless, it remains an open question whether such delocalization can give rise to diffusive transport in flat-band systems, and how the diffusion coefficient is related to the underlying quantum geometry.

In this paper, we address this open question by investigating disorder-driven quantum transport in flat-band systems with nontrivial quantum geometry. Using the Landauer-B\"uttiker formalism~\cite{Landauer-1D,Buttiker-1,Buttiker-2,FourTermMeasure-2,FourTermMeasure1,Multiterminal-1}, we study a four-terminal metal/flat-band/metal (M/FB/M) junction based on diamond lattice. By measuring the transmission $\mathcal{T}$ between two central leads in the presence of disorder, we reveal that disorder-induced diffusive transport in isolated flat bands can be characterized by quantum geometry. In the clean limit, transport is mediated solely by interface-bound states whose localization length is set by the quantum geometry of Bloch waves. Remarkably, disorder enables bulk-state transmission at zero energy, leading to a pronounced enhancement of transport. We further confirm, via wave-packet dynamics, that this delocalized diffusive transport is governed by the quantum metric. Finally, we provide a theoretical derivation showing that disorder generates an effective nonzero velocity operator, proportional to both disorder strength and quantum metric, thus establishing a direct link between disorder-induced diffusive delocalization and quantum geometry in flat-band.


\emph{\color{blue}M/FB/M junction.—}\label{diamond} The M/FB/M junction is constructed by connecting a central diamond lattice to two metallic leads, as depicted in Fig.~\ref{Fig:junction}(a). Each unit cell of the diamond lattice hosts three orbitals (A, B, and C), with corresponding annihilation operators $\hat{a}_{x}$, $\hat{b}_{x}$, and $\hat{c}_{x}$. The Hamiltonian for diamond lattice reads $\hat{H}_{\text{diamond}} = \sum_x \hat h_x$ with
\begin{align}
   \!\!\!\! \hat h_{x} &=J_{+}(\hat{b}_{x}^{\dagger}\hat{a}_{x}+\hat{c}_{x}^{\dagger}\hat{a}_{x})+J_{-}(\hat{a}_{x}^{\dagger}\hat{b}_{x+1}+\hat{c}_{x}^{\dagger}\hat{a}_{x+1})+h.c.,
\end{align}
where $J_{\pm} = J(1 \pm \delta)$ are the intra- and inter-cell hopping amplitudes respectively, with $x$ labeling the unit cell. In our calculations, a chemical potential is also introduced in the middle region to simulate gating. The diamond lattice features a flat band separated from two dispersive bands by a gap $\Delta = 2\sqrt{2}J\delta$ as illustrated in Fig.~\ref{Fig:junction}(b).
The quantum metric for a Bloch state $\vket{u(k)}$ is defined as
\begin{align}
\mathcal{G}(k) & = \langle\partial_{k}u(k)|(1-|u(k)\rangle\langle u(k)|)|\partial_{k}u(k)\rangle,
\end{align}
with its Brillouin-zone average for the flat-band state $|u_0(k)\rangle$ given by
\begin{equation}\label{eq:qm}
    \overline{\mathcal{G}}=\frac{a}{2\pi}\int_{-\pi/a}^{\pi/a}\mathcal{G}(k)dk=\frac{a^2(1-\delta)^2}{8\delta},
\end{equation}
where $a$ is the lattice constant, and we set $a=1$ throughout this work. Previous studies~\cite{QM_SC_1,QM_SC_2,Maj_1,FB-JJ-3} have shown that the quantum metric length in Eq.~\eqref{eq:qm} can provide a characteristic length scale for the underlying physics.

To minimize finite-size effects, the central diamond lattice of length $L$ is connected at both ends to identical clean diamond lattice leads (leads 1 and 4), effectively forming an infinite chain [Fig.~\ref{Fig:junction}(c)]. Two additional metallic leads (2 and 3) with hopping $t$ are coupled to the central region with coupling strength $T_\partial$ to probe the transmission $\mathcal{T}$. Disorder is introduced only in the central diamond lattice, while all leads remain clean. Since the clean flat band does not support bulk transport, transmission between leads 1 and 4 vanishes; thus, we focus on the transmission between leads 2 and 3, with the relevant device length given by $L_{23}$.

\begin{figure}[t]
\includegraphics[width=1\columnwidth]{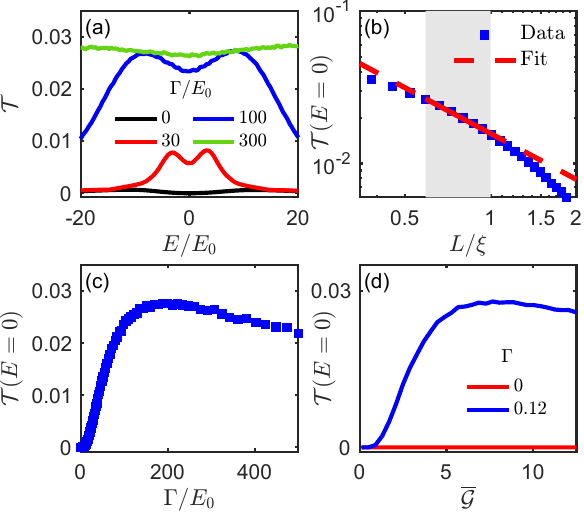}
\caption{(a) Transmission profiles for varying disorder strength $\Gamma$ at $\delta=0.01$ for a $L=50$ junction. As we increase the disorder strength, transport from bulk states is gradually activated. No zero energy transmission is observed in the clean limit. (b) The transmission $\mathcal{T_0}$ at zero energy $E=0$ and $\mathcal{T}\sim1/L$ fit (red line) for varying junction length $L$ when $\delta=0.01$ and $\Gamma=300E_0$. The gray shaded region indicates the diffusive $1/L$ region. The $\xi$ is chosen as the length when diffusive behavior holds.  (c) The transmissions $\mathcal{T}(E=0)$ for different disorder strength at $\delta=0.01$. Transmission contributed from bound states dominates when $\Gamma/E_{0}(\delta)$ is small, and $\mathcal{T}(E=0)$ increases as $\propto\Gamma^{2}$. A further increase in disorder strength enhances transport, peaking at $\Gamma/E_{0}(\delta)\sim200$. (d) The zero energy transmission $\mathcal{T}(E=0)$ for different $\overline{\mathcal{G}}$ at clean limit and a fixed disorder strength $\Gamma=300E_0(\delta=0.01)=0.12$ with $L=50$.
}
\label{Fig:4T}
\end{figure}


\emph{\color{blue}Disorder-free case.—} As shown in Fig.~\ref{Fig:4T}(a), there is no zero energy transmission in the absence of disorder. Rather, in the clean limit, when the two metallic leads are coupled to the flat band of diamond lattice, two interface bound states can be formed with the decay length being tuned by quantum metric~\cite{supple}. The two interface state, originating from the right (left) interface, has a decay length $\lambda = 1/2\delta$. When $\lambda$ is comparable to the junction length, the two interface states hybridize, as such the finite overlap constitute a weak channel for particles to tunnel as their energy deviates from the zero energy flat band, thus creating two separate peaks in the transmission profile. However, this hybridization does not initiate any direct current(DC) transport at $E=0$. This is because of the large degeneracy of bound states as long as the wavefunction $\psi_A$ at sublattice A vanishes, which are unstable and can easily be smeared by scattering. Thus, bulk-state transmission is absent in the clean limit and finite transmission requires bound states with nonzero energy.

By solving the full wave functions for the two-terminal case~\cite{supple}, for $|E| \ll \Delta$, the wavefunction at sublattice A is
\begin{equation}
\psi_{\text{A}}(x) = \frac{(-1)^{x}}{\sqrt{2}} \frac{E}{\Delta} \left[ b_0 e^{\delta(1+2x)} + c_0 e^{\delta(1-2x)} \right].
\end{equation}
The boundary conditions, set by the leads, determine $b_{0}$ and $c_{0}$. Including the effect of the leads, the transmission from interface-bound states is~\cite{supple}
\begin{equation}
\mathcal{T}^{-1}(E) = 1 + \left[ (1-\kappa^2) \frac{E^2 + E_0^2}{4\kappa E E_0} \right]^2,
\end{equation}
for $|E|\ll t$, where $\kappa=e^{-L/\lambda}$ is the exponential decay factor with decay length $\lambda$. For small disorder strength $\Gamma/E_{0}(\delta)<10$, $\mathcal{T}(E)$ exhibits two peaks at $\pm E_{0}(\delta)\approx\pm4T_{\partial}^{2}\delta/t$ [Fig.~\ref{Fig:4T}(a)].

We verify our theoretical predictions for the transmission profile through transport measurements. In the junction setup, coupling between the metallic leads and the flat-band material yields a characteristic peak energy, $E_{0}(\delta)$~\cite{supple}. This peak energy, together with the maximum transmittance, defines the observed transmission profile, as shown in Fig.~\ref{Fig:4T}(a). In the weak disorder limit, the transmission between leads 2 and 3 exhibits a double-peak structure at $E \sim \pm E_0(\delta)$, arising from the hybridization of metal–flat band interface bound states. In particular, we highlight that the zero energy conductance remains zero, reflecting the localization of flat-band states, in agreement with previsous theoretical expectations~\cite{Kubo-Greenwood-1,FB-transport-2}.

\begin{figure}[t]
\includegraphics[width=1\columnwidth]{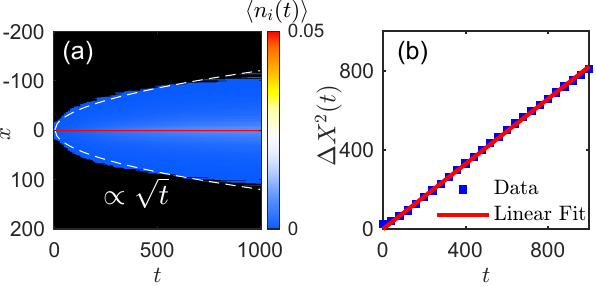} 
\caption{(a) The time evolution of the site occupation $\average{n_i(t)}=\average{\sum_\alpha |\psi_{i\alpha}|^2}$ for the wave packet $\vket{\psi(t)}$ formed by flat band states. 
 (b) The linear fit of the MSD $\Delta X^2(t)=2Dt$ with $D\approx 0.4109$. The fitted diffusion coefficient $D$ is close to 
the one value $0.4213$ in Table.~\ref{tab:tab1} predicted from Eq.~\eqref{eq:diffusion}. 
The parabolic behavior at the beginning part may contribute to the disorder-free region around the initial wave packet such that it can propagate ballistically shortly.
The evolution is performed on 1D diamond lattice with the length $L=401$ under the parameters  $\Gamma=0.1, \delta=0.01$ by averaging over $500$ disorder realizations. 
}
\label{Fig:Diffusion} 
\end{figure}


\emph{\color{blue}Numerics on disorder effects.---} Disorder can break the quantum interference underpinning the localization of flat-band states. To activate bulk transport, here we introduce Anderson-type onsite disorder to the diamond lattice:
\begin{equation}
\hat{H}_{\text{dis}}=\sum_{x}w_{x}(\hat{a}_{x}^{\dagger}\hat{a}_{x}+\hat{b}_{x}^{\dagger}\hat{b}_{x}+\hat{c}_{x}^{\dagger}\hat{c}_{x}),\label{eq:dis}
\end{equation}
where the onsite disorder $w_{x}$ is independently and uniformly distributed in $[-\Gamma/2, \Gamma/2]$. As shown in Fig.~\ref{Fig:4T}(a), for weak disorder ($\Gamma<E_{0}(\delta)$), the transmission exhibits two peaks at $E=\pm E_{0}(\delta)$. As the disorder strength increases ($\Gamma\gg E_{0}(\delta)$), these peaks broaden and merge into a plaquette-like structure, indicating the suppression of interface states and the emergence of bulk-state transmission. Since the flat-band states are initially localized and the transmission is not quantized as in conventional one-dimensional channels, the resulting bulk transport is inherently non-ballistic for sufficiently long junctions.

To understand the bulk-state transport in the presence of disorder, we examine the dependence of the transmission on the sample size. We compute the zero-energy transmission $\mathcal{T}(E=0)$ as a function of junction length $L$ in Fig.~\ref{Fig:4T}(b). Three distinct transport regimes are observed, separated by the localization length $\xi$, which is consistent with the scaling law~\cite{Dis-Cond-1}. In the ballistic regime ($L \ll \xi$), the transmission decreases linearly with length, following $\mathcal{T} \propto 1 - L/\xi$~\citep{Q1D-Diffu}; here, scattering is minimal and transport remains nearly ballistic. As the junction length approaches the localization length ($L \sim \xi$), the system enters the diffusive regime, characterized by Ohmic scaling ~\cite{Diff-Len-1} $\mathcal{T} \propto L^{-1}$, evident by a straight line of slope $-1$ in the log-log plot of Fig.~\ref{Fig:4T}(b). In this regime, disorder broadens the bandwidth of the flat band and disrupt the quantum interference which leads to localization, allowing the localized electrons to propagate with obstructions as in conventional disordered systems~\citep{Review-QuantTrans}. To enable a diffusive transport, we need to answer the origin of the finite group velocity, which will be discussed latter. For sufficiently long junctions ($L \gg \xi$), the system transitions to the localized regime, where Anderson localization dominates and transmission decays exponentially with length, $\mathcal{T} \propto e^{-L/\xi}$.

We also calculated the influence of disorder strength on transmission in Fig.~\ref{Fig:4T}(c). When the system has no disorder, the zero energy transmission is fixed at 0 and has no transport. As we slightly increase disorder strength, $\mathcal{T}(E=0)$ increases with $\Gamma/E_{0}(\delta)$, indicating the delocalization effect of disorder until reaching a maximum at $\Gamma/E_{0}(\delta)\sim200$. For stronger disorder, zero-energy transmission is governed by bulk transport and decreases with increasing $\Gamma$~\citep{Dis-Cond-1}. This decreasing transmission behavior, distinct from the enhanced interface transport, is consistent with conventional conductance in one-dimensional disordered systems.

As the Wannier function may extend over the bulk, the quantum metric which measures the overlap of Wannier wave functions\cite{QM-Cond-Mat,Max-Localize-1,Max-Localize-2,FB_2} can also influence the bulk transport. According to \eqref{eq:qm}, we can vary the quantum metric $\overline{\mathcal{G}}$ by tuning $\delta$. In Fig.~\ref{Fig:4T}(d), we present the zero-energy transmission $\mathcal{T}(E=0)$ for different $\overline{\mathcal{G}}$ in both clean and disordered regimes while keeping the gap $\Delta$ fixed. In the clean system, the destructive interference localize the flat band states and suppresses DC transmission even with a large quantum metric. Upon introducing disorder, the DC transport is initiated. As $\overline{\mathcal{G}}$ increases, the zero-energy transmission is enhanced, since the increase of quantum metric will increase the overlap of Wannier wave function, with disorder disrupting the interference effect, the electron's hopping will become stronger as the $\overline{\mathcal{G}}$ becomes larger~\citep{QM-Addedcontent}. 


\emph{\color{blue} Wave packet dynamics.—} To further confirm the diffusive transport observed in the M/FB/M junction setup, we can study the short-time behavior through the wave packet dynamics to overcome finite size effect. In the wave packet dynamics, the diffusion coefficient can be extracted from the time dependent mean square displacement (MSD) $\Delta X^2(t)$ as $D=\frac{1}{2}\frac{d(\Delta X^2)}{dt}$, which can be calculated through ~\cite{Diffusion-MSD-1,Diffusion-MSD-2,FB-WavePacket-1,FB-Photon-1}:
\begin{equation}\label{eq:MSD}
\Delta X^2(t)=\sum_{i=-L/2}^{L/2}i^2 \average{n_i(t)}-\left(\sum_{i=-L/2}^{L/2}i\, \average{n_i(t)}\right)^2,
\end{equation}
where $\average{\cdot}$ is the disorder average and $n_i(t)=\sum_\alpha |\psi_{i\alpha}(t)|^2$ is the occupation number at site $i$ at time $t$. The MSD $\Delta X^2(t)$ is measuring of how far the wave packet has spread over time. In particular, if the wave packet evolves diffusively, the MSD will grow linearly with time, $\Delta X^2(t)=2Dt$.

In Fig.~\ref{Fig:Diffusion}, we initialize a wave packet composed of disorder-free flat-band states, which is essential to reveal diffusive transport masked by the ballistic transport of dispersive bands~\cite{DL-Diffusion}. Then we turn on the disorder at $t=0$ and evolve the system under the perturbed Hamiltonian~\cite{supple}. As shown in Fig.~\ref{Fig:Diffusion}(b), the MSD exhibits a linear dependence on time, indicating that an initially localized wave packet diffuses via a random walk process~\cite{Diffusion-MSD-1,Diffusion-MSD-3,Diffusion-Noise-1} when disorder is present. To understand the diffusive transport observed, we have to address two questions: what sets the diffusion length, and how does disorder give rise to a finite group velocity in flat band?

\begin{table}[t]
\caption{Diffusion coefficients $D_\text{pred}$ calculated through Eq.~\eqref{eq:diffusion} and $D_\text{numeric}$ obtained from numerical fitting of Eq.~\eqref{eq:MSD}  with an example shown in Fig.~\ref{Fig:Diffusion}(b). The system parameters are listed, and all data are computed for a chain of length $L=1001$, averaged over 20 disorder realizations.}
\begin{ruledtabular}
\begin{tabular}{ccccc}\label{tab:tab1}
  $J$& $\delta$ & $\Gamma$ & $D_\text{pred}$ & $D_\text{numeric}$ \\ 
  \midrule
  1000 & 0.10 & 0.10 & $0.0421$ & $0.0182$\\ 
  1000 & 0.10 & 0.01 & $0.0042$ & $0.0026$\\ 
  1000 & 0.01 & 0.10  & $0.4213$ & $0.4338$ \\ 
  1000 & 0.05 & 0.07  & $0.0590$ & $0.0442$ \\ 
  10000 & 0.01 & 0.10  & $0.4213$ & $0.3744$ \\ 
  100000 & 0.01 & 0.10  & $0.4213$ & $0.4184$ \\ 
  100000 & 0.03 & 0.20  & $0.2808$ & $0.2493$ \\ 
\end{tabular}
\end{ruledtabular}
\end{table}


\emph{\color{blue} Diffusion in flat band.—} The introduction of disorder breaks the quantum interference underpinning the compact localization of flat-band states~\citep{CLS-1,CLS-2}. To understand the dependence of transmission on disorder, we use the decay length as the characteristic transport scale, anticipating that bound states can be excited by disorder. Assuming the retarded Green function for the flat band system is $G=1/(E+i\eta)$, where $\eta\rightarrow 0^+$, introduction of disorder leads to broadening of the flat band, and the disorder-averaged Green function is given by $\overline{G}(E)=1/(E+i\Gamma)$ for $|E|<\Gamma$, and otherwise $\overline{G}(E)=1/E$. For weak disorder $\Gamma<E_{0}(\delta)$, where the leading contribution is from a single scattering process, the transmission is given by $\mathcal{T}(E)=16e^{-4L\delta}\Gamma^{2}E_{0}^{2}(\delta)/(\Gamma+E_{0}(\delta))^{4}$ for $|E|\ll\Gamma$~\citep{supple}. Thus, the broadening of the interface bound state transmission profile enhances zero-energy transport.

A central question in mesoscopic physics is identifying a characteristic length scale that governs diffusion in flat-band systems. This diffusion length can be derived from the density-density correlation function~\cite{supple,Book-Meso}, restricted to the flat-band subspace. We focus on the intraband contributions from the flat band and employ the ladder approximation, where the impurity vertex $\Pi(\omega,q)$ satisfies the Bethe-Salpeter equation, describing the diffuson process~\citep{Book-Meso}:
\begin{equation}\label{eq:BetheSalpeter}
\Pi(\omega,q)=\Pi_{0}(\omega,q)+P_{0,\omega}\Pi_{0}(\omega,q)\Pi(\omega,q),
\end{equation}
with the bare impurity vertex $\Pi_{0}(\omega,q)=\int\frac{dk}{2\pi}|\langle u(k)|u(k+q)\rangle|^{2}$ and the quantum diffusion probability without collisions $P_{0,\omega}=\overline{G}(E)\overline{G}(E+\omega)$. In the small $q$ limit, $\Pi_{0}(\omega,q)\approx\Gamma^2(1-q^{2}\overline{\mathcal{G}})$, where $\overline{\mathcal{G}}=\int\frac{dk}{2\pi}\mathcal{G}(k)$ is the quantum metric averaged over the Brillouin zone. Solving Eq.~\eqref{eq:BetheSalpeter}, we obtain the diffusion coefficient to lowest order~\citep{supple}:
\begin{equation}\label{eq:diffusion}
    D=C\times\Gamma\, \overline{\mathcal{G}},
\end{equation}
revealing that the quantum metric $\overline{\mathcal{G}}$ sets the characteristic diffusion length in flat-band systems. Numerical simulations, detailed in the Supplemental Material~\cite{supple}, yield a proportionality constant $C\approx 0.337$. Table.~\ref{tab:tab1} presents our wavepacket simulation results for various parameters and corresponding diffusion coefficients. These results show that our estimates of the diffusion coefficient, based on Eq.~\eqref{eq:diffusion}, agree well with those obtained through MSD fitting, especially for small $\delta$. Additionally, the diffusion coefficient is robust against changes of the hopping strength $J$. Our result also resembles the coherence length from the quantum metric in a flat-band superconductor. Diffuson can be associated with particle-hole excitations, thus is analogous to a Cooper pair in a flat-band superconductor~\citep{QM_SC_1,QM_SC_2}, suggesting that the quantum metric naturally emerges as a characteristic length scale in such systems.

To enable finite zero-frequency transmission, a finite velocity operator is required according to the Kubo-Greenwood formula~\citep{Kubo-Greenwood-1,FB-Trans-1}, $\mathcal{T}\sim\overline{{\rm Tr}[\Im G(E)\hat{v}\Im G(E)\hat{v}]}$. We approximate $\mathcal{T}\sim{\rm Tr}[\Im\overline{G}\overline{\hat{v}}\Im\overline{G}\overline{\hat{v}}]$, where $\overline{\hat{v}}$ is the disorder-averaged velocity operator. The flat band is broadened by disorder, yielding finite $\Im\overline{G}$, which is maximal when $E=0$. Thus, to obtain finite DC transport, $\overline{\hat{v}}$ must be finite. In the band basis, the velocity operator is
\begin{align}
\!\! v_{nm}(k)=(\epsilon_{n}(k)-\epsilon_{m}(k))\langle \partial_{k}u_{n,k}|u_{m,k}\rangle+\partial_{k}\epsilon_{n}(k)\delta_{nm}.
\end{align}
Note the interband velocity operator is proportional to the band gap. The disorder term $\hat{H}_{\mathrm{dis}}=\sum_{kq}\sum_{mn}\frac{w_{q}}{V_{0}}\Gamma_{mn}(k,q)\hat{c}_{mk}^{\dagger}\hat{c}_{nk-q}$ can drive interband hopping with form factor $\Gamma_{mn}(k,q)=\langle u_{m,k}|u_{n,k+q}\rangle$. Thus, a correction of order $O(1)$ arises from the interplay between interband velocity operator and disorder. Diagrammatic expansion shows the leading order comes from a single disorder scattering, with the vertex being the interband velocity $\hat{v}_{0n}$, where $0$ and $n$ denote the flat band and bands, respectively. Thus, for $|E|<\Gamma$,
\begin{equation}
\overline{v}_{00}(k)\propto2\Gamma\int\frac{dq}{2\pi}\mathrm{Re}\langle u_{0,k}|\partial_{k}u_{0,k+q}\rangle\langle u_{0,k+q}|u_{0,k}\rangle,
\end{equation}
which is proportional to disorder strength, with the sum over $q$ arises from disorder scattering. With both the diffusion coefficient and effective velocity, diffusive transport can contribute to the zero conductivity absent in the clean limit. Using the Einstein relation, we estimate the conductance as $\mathcal{T}\sim D\rho(E)/L\sim 0.04$, close to the value shown in Fig.~\ref{Fig:4T}(a).


\emph{\blue{Discussion.}---}  The results presented above allow us to explore disorder induced delocalization in flat band systems with quantum geometry. In non-interacting flat-band systems, the spatial spread of Wannier functions is governed by the quantum geometry of the flat band\citep{FB_2}. In a finite-sized system, transmission is influenced by the spread of Wannier functions at the system’s interface, while bulk states remain localized due to destructive interference. However, the introduction of onsite disorder distorts this perfect destructive interference, enabling localized particles to hop and acquire an effective velocity. When the disorder strength is sufficiently weak to prevent the connection with dispersive bands but strong enough to deviate states from the flat band, a wave packet composed of flat-band states diffuses with obstructions, resembling multiple scattering events. This wiggling evolution of wave packet evolution can be interpreted as diffusive behavior, leading to the delocalization of flat-band states.

As disorder strength increases further, the system transitions out of the flat-band localization regime, and disorder begins to suppress wave propagation, signaling the re-entrance of  localization, specifically Anderson localization. This transition has been experimentally verified in the one-dimensional Tasaki lattice\citep{And-Loc-1,FBL-AL-1D-Exper} and in superconducting qubit array \citep{FB-SC-Qbit-Array}. In the Tasaki lattice, subtle signatures of particle population diffusion are observable when the band is tuned to be flat. However the absence of quantum metric in Tasaki lattice\citep{Flat-Tasaki-1} and the imperfect interatomic interactions may obscure disorder-induced diffusive wave packet behavior in flat bands. In contrast, we expect that the diamond lattice, with its isolated flat band and tunable quantum metric, should exhibit more pronounced experimental evidence of diffusion.


\emph{\color{blue}Conclusion.—} Flat-band materials such as moiré patterns~\citep{Moire_1,Moire_2,Moire_3}, Kagome lattices~\citep{Kagome_1}, artificial quantum dot arrays~\citep{QDA}, or optical lattices~\citep{Op_Lat_1} could be used to construct M/FB/M junctions. The quantum geometry can be tuned by adjusting parameters such as twist angle or lattice geometry, making these materials promising for realizing the M/FB/M junction concept. Such experiments would not only validate our theoretical predictions but also pave the way for novel quantum devices exploiting the unique transport properties of flat-band systems. Our numerics show that disorder does not suppress transport in flat-band systems, but instead enhances it, shedding light on why realistic flat-band systems—such as twisted bilayer graphene, where disorder is intrinsic—still exhibit robust transport at low carrier density.

\begingroup
\renewcommand{\addcontentsline}[3]{}
\renewcommand{\section}[2]{}
\begin{acknowledgments} 
	We thank Patrick A. Lee, Tai-Kai Ng, Roderich Moessner, Akito Daido, Sen Mu, and Haijing Zhang for their valuable discussions. K. T. L. acknowledges the support of the Ministry of Science and Technology, China, and the Hong Kong Research Grants Council through Grants No. 2020YFA0309600, No. RFS2021-6S03, No. C6025-19G, No. AoE/P-701/20, No. 16310520, No. 16310219, No. 16307622, and No. 16309223. 
\end{acknowledgments}

%

\endgroup


\clearpage
\onecolumngrid

\begin{center}
	\textbf{\large Supplemental Material for ``Disorder-induced Diffusion Transport in Flat-band Systems with Quantum Metric"}\\[.2cm]
	Chun Wang Chau$^{*1,3}$, Tian Xiang$^{*1}$, Shuai A. Chen$^{\dagger1,2}$, K. T. Law$^{\ddagger1}$\\[.1cm]
	{\itshape 1. Department of Physics, Hong Kong University of Science and Technology, Clear Water Water Bay, 999077 Hong Kong, China}\\
	{\itshape 2. Max Planck Institute for the Physics of Complex Systems, N\"{o}thnitzer Stra{\ss}e 38, Dresden 01187, Germany}\\
	{\itshape 3. Cavendish Laboratory, Department of Physics, J J Thomson Avenue, Cambridge CB3 0HE, United Kingdom}\\[1cm]
\end{center}

\maketitle
\def\intinf{\int_{-\infty}^{\infty}}
\def\dm{\mathrm{d}}
\def\trace{{\rm Tr}}
\def\ubar#1{\underline{#1}}
\def\Re{\mathbf{Re}}
\def\Im{\mathbf{Im}}
\providecommand{\bra}[1]{\langle#1}
\providecommand{\ket}[1]{#1\rangle}
\providecommand{\vbra}[1]{\langle#1\rvert}
\providecommand{\vket}[1]{\lvert#1\rangle}
\providecommand{\average}[1]{\langle#1\rangle}
\renewcommand{\thefigure}{S\arabic{figure}}
\renewcommand{\theequation}{S\arabic{equation}}

\newcommand{\pd}[2]{\frac{\partial #1}{\partial #2}}
\newcommand{\bs}[1]{\boldsymbol{#1}}
\newcommand{\reals}{\mathbb{R}}
\newcommand{\complex}{\mathbb{C}}
\newcommand{\abs}[1]{\left\lvert#1\right\rvert}

\makeatother

\setcounter{equation}{0}
\setcounter{section}{0}
\setcounter{figure}{0}
\setcounter{table}{0}
\setcounter{page}{1}
\renewcommand{\theequation}{S\arabic{equation}}
\renewcommand{\thesection}{S\Roman{section}}
\renewcommand{\thefigure}{S\arabic{figure}}
\renewcommand{\thetable}{\arabic{table}}
\renewcommand{\tablename}{Supplementary Table}

\renewcommand{\bibnumfmt}[1]{[S#1]}
\renewcommand{\citenumfont}[1]{S#1}


\setcounter{secnumdepth}{2}
\tableofcontents

\section{Quantum geometry of diamond lattice}

\label{Quantum_Metric} Recall the Bloch state of flat band  with lattice constant $a$ is given
by 
\begin{equation}
	\vert u_{0}(k)\rangle=\frac{1}{\epsilon(k)}\begin{pmatrix}0\\
		e^{-ika}J_{-}+J_{+}\\
		-(e^{ika}J_{-}+J_{+})
	\end{pmatrix},\label{eq:S-u0}
\end{equation}
with $\epsilon(k)=\sqrt{2(J_{+}^{2}+J_{-}^{2}+2J_{+}J_{-}\cos{ka})}$ being the dispersion relation of dispersive bands. This gives out the band gap $\Delta=\epsilon(k=\pi/a)=2\sqrt{2}J\delta$. 

As such the quantum metric is given by: 
\begin{align}
	\mathcal{G}(k) & =\mathrm{Re}\langle\partial_{k}u_{0}(k)|(1-|u_{0}(k)\rangle\langle u_{0}(k)|)|\partial_{k}u_{0}(k)\rangle\nonumber \\
	& =\langle\partial_{k}u_{0}(k)|\partial_{k}u_{0}(k)\rangle\nonumber \\
	& =\frac{a^2(1-\delta)^2[1-\delta+(1+\delta)\cos ka]^2}{4[1+\delta^2+(1-\delta^2)\cos ka]^2}
\end{align}
With the average given by: 
\begin{align}
	\bar{\mathcal{G}} & =\frac{1}{2\pi}\int_{-\pi/a}^{\pi/a}dk\ \mathcal{G}(k)\nonumber \\
	& =\frac{a(1-\delta)^{2}}{8\delta}.
\end{align}
If we take $a=1$, for $\delta\ll1$, $\bar{g}=1/8\delta$, which is one-fourth of the
decay length of the interface states we had discussed in the maintext. 

\section{Interface state wave function and the transmission}

\label{Wavefunction}

\begin{figure}[t]
	\centering \includegraphics[width=0.7\columnwidth]{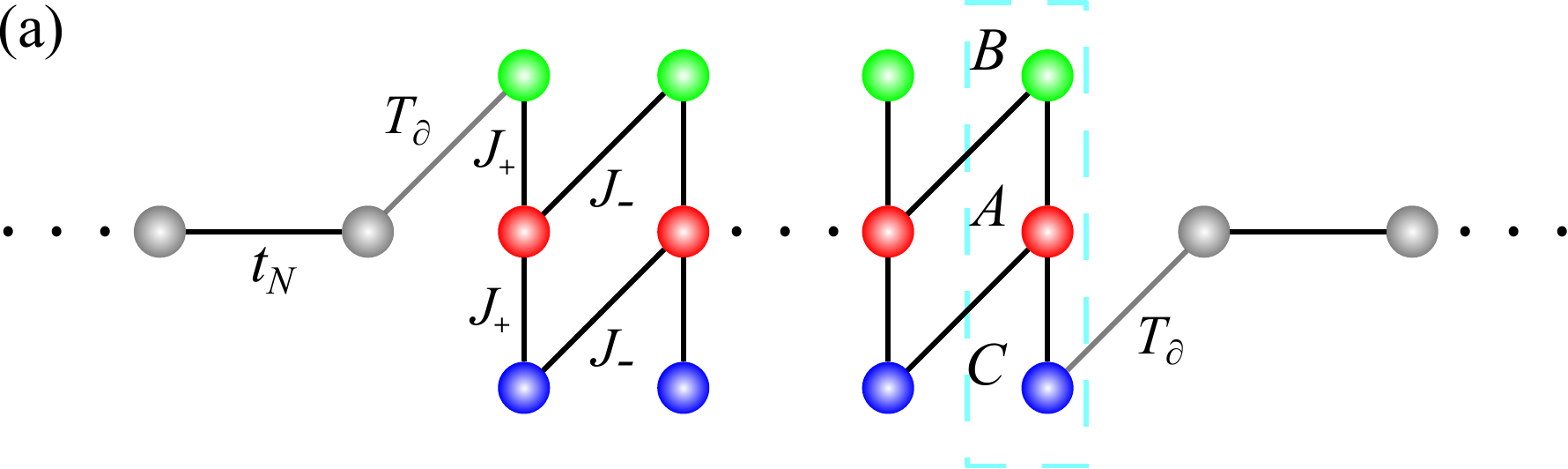} \includegraphics[width=0.3\columnwidth]{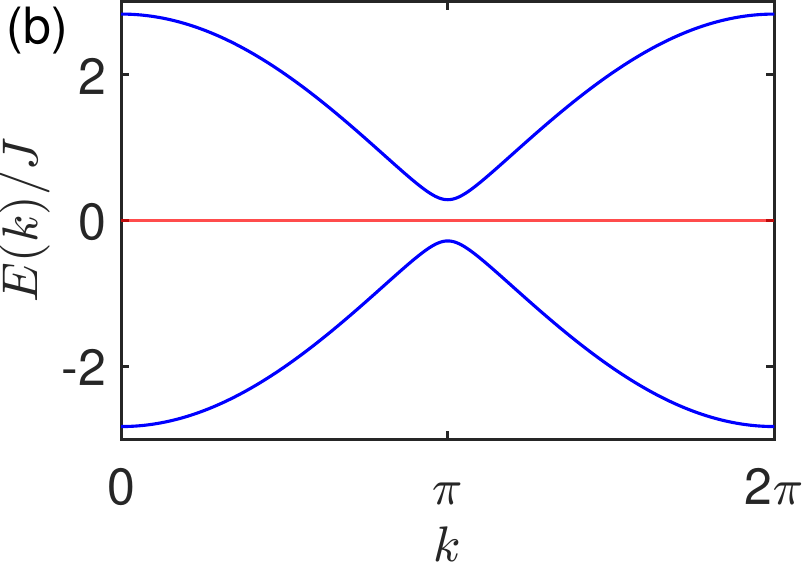}
	\includegraphics[width=0.3\columnwidth]{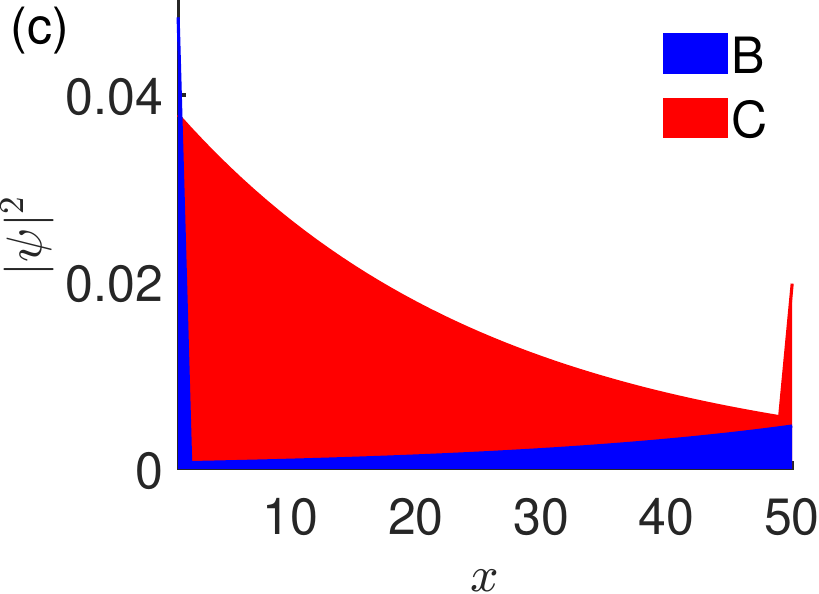} \caption{ (a) Structure of M/FB/M junction. The flat band originates from the
		diamond lattice, which has three lattice sites $A,B,C$ per unit cell.
		(b) Dispersion spectrum of the diamond lattice for $\delta=0.1$. (c)
		Density distribution of a pair of interface states within the diamond
		lattice of M/FB/M junction with parameter $\delta=0.01$ and length
		$L=50$. The two interface states are located at the B- and C- sublattice
		sites, respectively, with a localization length $\xi=1/(2\delta)$.}
	\label{Fig:S-junction} 
\end{figure}

In this section, we give details on the derivation of the bound state
wave functions. Given the setup of the M/FB/M junction in Fig.~\ref{Fig:S-junction}(a),
the bound state energy is determined by the incoming wave. At zero
energy $E=0$, the flat dispersion allows us to do the linear combinations
of scattering states to a bound state. Write down the Hamiltonian near
the lead for the flat band 
\begin{equation}
	\left(\begin{array}{c|cc|ccccccc}
		\ddots & -1 & 0 & a_{1} & b_{1} & c_{1} & a_{2} & b_{2} & c_{2} & \cdots\\
		\hline -1 & 0 & t_{N} & 0 & 0 & 0 & 0 & 0 & 0 & \cdots\\
		0 & t_{N} & 0 & 0 & T_{\partial} & 0 & 0 & 0 & 0 & \cdots\\
		\hline a_{1} & 0 & 0 & 0 & J_{+} & J_{+} & 0 & J_{-} & 0 & \cdots\\
		b_{1} & 0 & T_{\partial} & J_{+} & 0 & 0 & 0 & 0 & 0 & \cdots\\
		c_{1} & 0 & 0 & J_{+} & 0 & 0 & J_{-} & 0 & 0 & \cdots\\
		a_{2} & 0 & 0 & 0 & 0 & J_{-} & 0 & J_{+} & J_{+} & \cdots\\
		b_{2} & 0 & 0 & J_{-} & 0 & 0 & J_{+} & 0 & 0 & \cdots\\
		c_{2} & 0 & 0 & 0 & 0 & 0 & J_{+} & 0 & 0 & \cdots
	\end{array}\right)\begin{pmatrix}\vdots\\
		\psi_{-1}^{(L)}\\
		\psi_{0}^{(L)}\\
		a_{1}\\
		b_{1}\\
		c_{1}\\
		a_{2}\\
		b_{2}\\
		c_{2}\\
		\vdots
	\end{pmatrix}=E_{flat}\begin{pmatrix}\vdots\\
		\psi_{-1}^{(L)}\\
		\psi_{0}^{(L)}\\
		a_{1}\\
		b_{1}\\
		c_{1}\\
		a_{2}\\
		b_{2}\\
		c_{2}\\
		\vdots
	\end{pmatrix}=\mathbf{0}.\label{eq:S-Hamiltonian}
\end{equation}
To simplify the notation, we denote $\alpha_{x}=\psi_{\alpha}(x)$
as the site wave function with $1<x<L$ denoting the unitcell of the
diamond lattice.

We can write down the secular equation for the wave function in the
bulk as 
\begin{equation}
	\begin{split} & J_{-}c_{x-1}+J_{+}b_{x}+J_{+}c_{x}+J_{-}b_{x+1}=0\\
		& J_{-}a_{x-1}+J_{+}a_{x}=0
	\end{split}
\end{equation}
From the structure of the Bloch wave in Eq.~\eqref{eq:S-u0}, the wavefunction on A
sublattice sites do not contribute, while the B and C sublattice sites
contribute equally, so the second equation become trivial while the first
equation can be reduced to a simpler form

\begin{align}
	0 & =J_{-}b_{x+1}+J_{+}b_{x},\label{eq:S-bxE0}\\
	0 & =J_{-}c_{x}+J_{+}c_{x+1},\label{eq:S-cxE0}
\end{align}

where we introduce the parameter $J_{+}=wJ$ and $J_{-}=w^{-1}J$.
From Eq.~\eqref{eq:S-bxE0}, we obtain the wave function 
\begin{equation}
	b_{x}=-(-1)^{x}w^{2x-2}b_{1},\label{eq:S-bx}
\end{equation}
with $b_{1}$ as the component at the left ending site. From Eq.~\eqref{eq:S-cxE0},
we have the solution 
\begin{equation}
	c_{x}=-(-1)^{x}w^{-2x-2}c_{1},\label{eq:S-cx}
\end{equation}
with $c_{1}$ as the component of the left ending site of C-sublattice.

By taking $w=1+\delta$ for small $\delta$, we retrieve the exponential
decay. This hints at, even if the energy is nonzero, these forms still
hold true for B and C sites up to a negligible perturbation. The key
difference is that wave function at A-sublattice sites is no longer
zero: 
\begin{align}
	Eb_{x} & =J_{+}a_{x}+J_{-}a_{x-1},\label{bx_a}\\
	Ec_{x-1} & =J_{+}a_{x-1}+J_{-}a_{x}.\label{cx_a}
\end{align}
As such one can obtain: 
\begin{align}
	a_{x} & =-\frac{E}{wJ}\frac{c_{x-1}-w^{2}b_{x}}{w^{2}-w^{-2}},\label{ax_1}\\
	& =-\frac{E}{wJ}\frac{b_{x+1}-w^{2}c_{x}}{w^{2}-w^{-2}}.\label{ax_2}
\end{align}
If we attact the 0th unit cell in the left, we can recover the result
in the eq(11) in the maintext 
\begin{equation}
	\psi_{A}(x)\equiv a_{x}=\frac{(-1)^{x}}{\sqrt{2}}\frac{E}{\Delta}\left[b_{0}e^{2\delta(x+1)}+c_{0}e^{-2\delta(x+1)}\right].\label{eq:S-psiA}
\end{equation}


At finite energy $E$, the degeneracy of the bound states is broken.
In Fig.~\ref{Fig:S-junction}(c), we plot a pair of interface bound
states given $E\neq0$ for the length $L=50$, and the bound states
are localized at two interfaces due to the coupling with the external
leads. For larger $E$ that is comparable with the band gap, our assumption
may not be valid since the dispersive band contributions to $\psi_{A}$
is no longer perturbative. Therefore, we reach the bound state solutions
for the diamond lattice within the M/FB/M junction as depicted by Eq.
\eqref{eq:S-psiA}. With the effect of the external leads included, we
can derive the transmission based on bound states.

We begin by considering the left end, where we have: 
\begin{align}
	a_{1} & =\frac{E}{wJ}\frac{w^{2}c_{1}-b_{2}}{w^{2}-w^{-2}}.\label{eq:S-a1_1}
\end{align}
Note that $a_{1}\ll b_{1},\ c_{1}$ given that $E\ll\delta J,\ t_{N}$.
As such we have: 
\begin{align}
	Ea_{1} & =J_{+}b_{1}+J_{+}c_{1}+J_{-}b_{2}\nonumber \\
	b_{1} & \sim-(b_{2}+c_{1}).\label{eq:S-b0}
\end{align}
For lead with chemical potential $\mu_{L}$, we note that: 
\begin{align}
	t_{N}e^{-ik_{L}}+t_{N}e^{ik_{L}} & =E-\mu_{L}\nonumber \\
	e^{ik_{L}} & =\frac{1}{2t_{N}}\left(E-\mu_{L}+i\sqrt{4t_{N}^{2}-(E-\mu_{L})^{2}}\right),
\end{align}
where we have assumed $0<k_{L}<\pi$. Taking the wave function on
left lead as $\psi_{x}^{(L)}=e^{ik_{L}x}+re^{-ik_{L}x}$, we have
boundary condition from lead in eq.\eqref{eq:S-Hamiltonian}: 
\begin{equation}
	(E-\mu_{L})\psi_{0}^{(L)}=t_{N}\psi_{-1}^{(L)}+T_{\partial}b_{1}
\end{equation}
\begin{equation}
	Eb_{1}=T_{\partial}\psi_{0}^{(L)}+J_{+}a_{1}
\end{equation}
for the first equation we can express $b_{1}$ in terms of $r$: 
\begin{align}
	(E-\mu_{L})(1+r) & =t_{N}(e^{-ik_{L}}+re^{ik_{L}})+T_{\partial}b_{1}\nonumber \\
	b_{1} & =\frac{(E-\mu_{L})(1+r)}{T_{\partial}}-\frac{t_{N}(e^{-ik_{L}}+re^{ik_{L}})}{T_{\partial}},\label{b0_1}
\end{align}
and for the second one we can use eq(\ref{eq:S-a1_1}) 
\begin{align}
	b_{1} & =\frac{T_{\partial}}{E}(1+r)+\frac{w^{2}c_{1}-b_{2}}{w^{2}-w^{-2}}\nonumber \\
	& \overset{\delta\ll1}{\sim}\frac{{T_{\partial}}}{E}(1+r)+\frac{c_{1}-b_{2}}{4\delta}\\
	4\delta b_{1}\sim0 & \sim\frac{4T_{\partial}\delta}{E}(1+r)+c_{1}-b_{2}\label{b0_2}
\end{align}

As such we can solve for $b_{2}$ and $c_{1}$ in terms of $r$ with
eq.\eqref{eq:S-b0}: 
\begin{align}
	b_{2} & \sim-\frac{1}{2}b_{1}+2\frac{T_{\partial}\delta}{E}(1+r),\\
	c_{1} & \sim-\frac{1}{2}b_{1}-2\frac{T_{\partial}\delta}{E}(1+r).
\end{align}
Similarly, if we assume there is only outgoing wavefunction $\psi_{x}^{(R)}=te^{ik_{R}x}$
on the right lead, we have: 
\begin{equation}
	\begin{split}a_{L} & =-\frac{E}{wJ}\frac{c_{L-1}-w^{2}b_{L}}{w^{2}-w^{-2}}\\
		Ea_{L} & =J_{-}c_{L-1}+J_{+}b_{L}+J_{+}c_{L}\\
		c_{L} & \sim-(b_{L}+c_{L-1})
	\end{split}
\end{equation}
and the right lead boundary 
\begin{equation}
	\begin{split}Ec_{L} & =J_{+}a_{L}+T_{\partial}\psi_{0}^{(R)}\\
		(E-\mu_{R})\psi_{0}^{(R)} & =T_{\partial}c_{L}+t_{N}\psi_{1}^{(R)}
	\end{split}
\end{equation}
which can be solved as 
\begin{align}
	b_{L} & \sim-\frac{c_{L}}{2}-2\frac{T_{\partial}\delta}{E}t,\\
	c_{L-1} & \sim-\frac{c_{L}}{2}+2\frac{T_{\partial}\delta}{E}t.
\end{align}
Where we have defined $2t_{N}\cos{k_{R}}=E-\mu_{R}$. Recall (\ref{eq:S-bx}-\ref{eq:S-cx}),
we have $b_{L}=\kappa^{-1}b_{2}$ and $c_{L-1}=\kappa c_{1}$, where
$\kappa\sim(-1)^{L-2}e^{-2\delta(L-2)}$. Considering only the linear
response, we can take $\mu_{L}=\mu_{R}=0$, which gives:

\begin{equation}
	t=\frac{i16E\delta\,\kappa\,T_{\partial}^{2}\,t_{N}\sin k}{(E^{2}(\kappa+1)-E(\kappa+1)e^{ik}t_{N}+4T_{\partial}^{2}\delta(\kappa-1))\left(E^{2}(\kappa-1)-E(\kappa-1)e^{ik}t_{N}+4T_{\partial}^{2}\delta(\kappa+1)\right)}
\end{equation}
For long enough junction, we have $\kappa\ll1$, which gives:

\begin{equation}
	t=-\frac{i16E\delta\kappa T_{\partial}^{2}t_{N}\sin k}{(Ee^{ik}t_{N}+4T_{\partial}^{2}\delta-\cancel{E^{2}})^{2}}.\label{eq:S-transcoeff}
\end{equation}
Omit the higher order term $O(E^{2})$ in the denominator, the transmission
coefficient $t$ can be related to the transmittance $\mathcal{T}$
as:

\begin{align}\label{eq:transmittance}
	\mathcal{T} & =\abs{t}^{2}\nonumber \\
	& \sim\frac{256E^{2}\delta^{2}\kappa^{2}T_{\partial}^{4}t_{N}^{2}\sin^{2}k}{(E^{2}t_{N}^{2}+\cancel{8\delta Et_{N}T_{\partial}^{2}\cos k}+16\delta^{2}T_{\partial}^{4})^{2}}\nonumber \\
	& \sim\frac{64\delta^{2}T_{\partial}^{4}}{t_{N}^{4}}\frac{E^{2}(4t_{N}^{2}-\cancel{E^{2}})}{(E^{2}+E_{0}^{2})^{2}}e^{-4\delta L}\nonumber \\
	& \sim\frac{16E^{2}E_{0}^{2}}{(E^{2}+E_{0}^{2})^{2}}e^{-4\delta L}.
\end{align}

Where we have used $2t_{N}\cos k=E$, $\kappa^{2}\sim e^{-4\delta L}$
and $E_{0}=4\delta T_{\partial}^{2}/t_{N}$. We have also assumed
that $t_{N}\gg E_{0}$. For junction of arbitrary length with $\delta\ll1$,
we have instead: 

\begin{equation}
	\mathcal{T}^{-1}=1+\left[(1-\kappa^{2})\frac{E^{2}+E_{0}^{2}}{4\kappa EE_{0}}\right]^{2}.\label{T_L}
\end{equation}

Where the maximum is still located at $E_{0}$, with maximal value
$\sim\mathrm{sech}^{2}(2L\delta)$. For reference, without detail
derivation, we note the most general form of transmittance is: 

\begin{equation}
	\mathcal{T}^{-1}=w^{8}+\left[\frac{E_{0}^{2}(\kappa^{2}-w^{4})+E^{2}(\kappa^{2}-w^{4}(1-2w^{4})^{2})}{4EE_{0}\kappa w^{2}}\right]^{2}
\end{equation}

\begin{figure}[t]
	\centering
	\includegraphics[height=0.355\columnwidth]{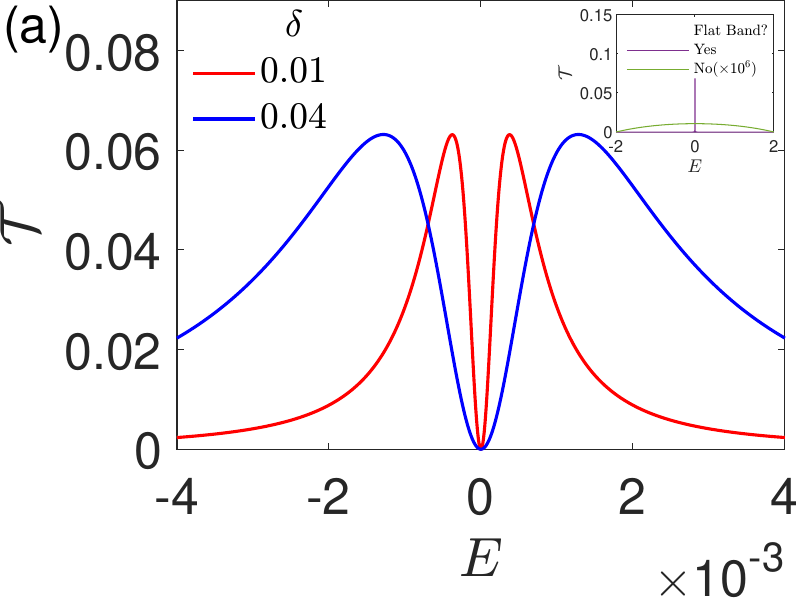} \includegraphics[height=0.35\columnwidth]{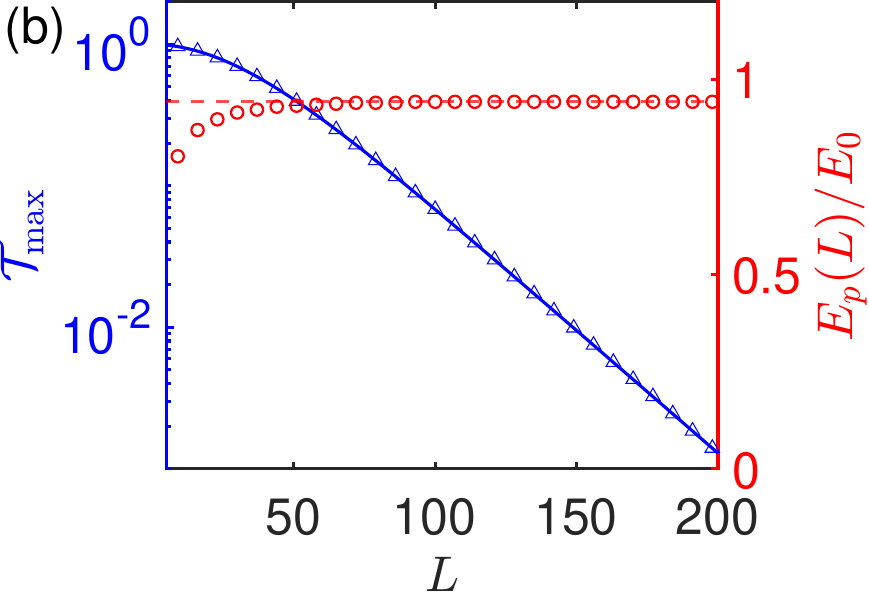}
	\caption{Two-terminal measurement on transmission in the clean limit: (a) transmission
		profile for different value of $\delta$, while keeping $L\delta\sim1$
		and (b) Maximal transmittance and peak energy as a function of the
		length of the lattice. We note the maximal transmittance is identical,
		with $E_{0}(\delta=0.04)\sim4E_{0}(\delta=0.01)$. In the inset, we
		compare the case of having and not having a flat band (2-band model
		with identical dispersive bands). We note the transmission is highly
		suppressed when the flat band is removed. In (b), when the junction
		is long enough, the peak energy $E_p(L)$ approaches a constant value $\sim E_{0}(\delta)$,
		and the maximal transmittance decays exponentially. In general, the
		maximal transmittance obeys $\mathcal{T}\sim\mathrm{sech}^{2}(2L\delta)$. }
	\label{fig:S-numeric} 
\end{figure}

Which reduces to \eqref{T_L} when $w\rightarrow1$, namely when $\delta\rightarrow0$.
This equation can explain the numerical result illustrated in Fig.~\ref{fig:S-numeric}(a-b)
exactly, but discussion was avoided in the main text due to non-trivial
function form.

Use the transmittance formula given above, for the short junction limit, we have the perfect transmission with
$\mathcal{T}_{\mathrm{max}}\rightarrow1$ at $\pm E_{0}$. On the
other hand, when the length of the junction is comparable to the localization
length, the transmittance can be simplified to 

\begin{equation}\label{Transmittance}
	\mathcal{T}(E)=\frac{16E^{2}E_{0}^{2}}{(E^{2}+E_{0}^{2})^{2}}e^{-4L\delta},
\end{equation}
which is maximal at $\pm E_{0}$ with $\mathcal{T}_{\mathrm{max}}=4e^{-4L\delta}$,
and recovers the case of the weak transmission limit of a square trap.
Eq.~\eqref{Transmittance} resembles the effect of a transport system
with two channels separated by energy $2E_{0}$. We can take $E_{0}$
as the characteristic energy scale for such an M/FB/M junction. In
Sec.~\ref{sec:S-Green}, we provide an alternate approach to derive the transmission within
the M/FB/M junction using Green's function method, which gives rise
to the same transmission profile as in Eq.~\eqref{Transmittance}
under the long junction limit.

As shown in Fig.~\ref{fig:S-numeric} the peak location is slightly smaller than the theoretical prediction $E_{0}(\delta)$. This discrepancy is due to the negligence of the higher-order terms in $\delta$ in the previous analytical calculation. However, the prediction remains valid when the junction is long enough ($L\delta>1$), where the peak location approaches a constant value close to $E_{0}(\delta)$.

To clarify the role of the flat band in transport, we can compare the transmission profile of the diamond lattice with the transmission profile of a two-band model without a flat band. The two-band model is constructed to contain the same dispersive bands as the diamond lattice except for the removal of the flat band. When the flat band is removed, the transmission is strongly suppressed by an order of $10^{-7}$ weaker, as shown in the inset of Fig.~\ref{fig:S-numeric}(a). Furthermore, the transmission profile reduces to the tunnel junction case with a single peak and a full width at half maximum (FWHM) on the order of $t_{N}$. Thus, we can conclude that the significant overall transmission as well as its two-peak profile is enabled by the flat band, where the small energy scale $E_{0}(\delta)$ emerges, allowing transmission to happen around the flat band.


\subsection{Degeneracy of flat band}

Below we will give a brief discussion on the degeneracy of the flat
band. In particular, we will limit our discussion to the transport
due to coupling of the interface state, instead of the propagating
state. By setting energy to zero, we have: 
\begin{align}
	0 & =wb_{1}+wc_{1}+w^{-1}b_{2}\\
	0 & =w\kappa^{-1}b_{2}+wc_{L}+w^{-1}\kappa c_{1}
\end{align}
Which gives: 
\begin{align}
	b_{1} & =-\frac{b_{2}+w^{2}c_{1}}{w^{2}}\\
	c_{L} & =-\frac{w^{2}b_{2}+\kappa^{2}c_{1}}{\kappa w^{2}}
\end{align}
Note that $b_{0}$ and $c_{L}$ can be directly related to reflectance
and transmittance, assuming there is either a $\pi$ or $0$ phase
shift upon reflection on the left boundary: 
\begin{equation}
	\frac{c_{L}}{b_{1}}=\frac{\sqrt{\mathcal{T}}}{1-\sqrt{\mathcal{R}}}
\end{equation}
Define $\alpha=b_{1}/c_{L}$, we have: 
\begin{align}
	\mathcal{T} & =\frac{4b_{1}^{2}c_{L}^{2}}{(b_{1}^{2}+c_{L}^{2})^{2}}\\
	& =\frac{4\alpha^{2}}{(1+\alpha^{2})^{2}}
\end{align}
Note that the range of the solution is always between $0$ and $1$,
fully transmitting when $\alpha=1$ and fully reflecting when $b_{1}=0$
and $c_{L}=0$. This demonstrate the degeneracy of the flat band,
and explain why no solution can be converged to numerically for the
transmittance at zero energy when we used the exact diagonization
approach.

\subsection{Born's approximation for disorder}

Using the Born's approximation and eq(\ref{eq:S-transcoeff}), the transmission
coefficient with disorder strength $\Gamma$ can be approximated by
substitution $E\to E+i\Gamma$: 
\begin{equation}
	t\sim-i\frac{16E\delta\kappa T_{\partial}^{2}t_{N}\sin k}{[(E+i\Gamma)e^{ik}t_{N}+4T_{\partial}^{2}\delta]^{2}}.
\end{equation}
In the limit where the junction is long enough $L\delta>1$, the transmittance
is given by: 
\begin{equation}
	\mathcal{T}\sim\frac{16(E^{2}+\Gamma^{2})E_{0}^{2}}{[E^{2}+(E_{0}+\Gamma)^{2}]^{2}}e^{-4\delta L},
\end{equation}
where the maximal value $\mathcal{T}_{\mathrm{max}}(\Gamma)=4e^{-4\delta L}E_{0}/(E_{0}+2\Gamma)$,
which decreases monotonically as we increase disorder. This is contrary
to our numerical calculation, where in the dirty limit with weak disorder
$\Gamma<E_{0}$, the transmission is enhanced instead of suppressed.
As such Born's approximation might not be a valid approach to consider
disorder within a flat band system.

The zero energy transmittance as a function of disorder is given by:
\begin{equation}
	\mathcal{T}(E=0)\sim\frac{16\Gamma^{2}E_{0}^{2}}{(E_{0}+\Gamma)^{4}}e^{-4\delta L}
\end{equation}
Recall at zero energy in clean limit transport is prohibited. As we
introduce disorder, for $\Gamma<E_{0}$, disorder enhances the transport,
until reaching a maximal at $\Gamma=E_{0}$ of transmittance $e^{-4\delta L}$.
The transmittance decrease as we further increase the disorder strength.

\section{Impurity pair calculation}

To demonstrate the effect of disorder in flat band system, below we
study the simplest case where correlation effect is important, namely
introducing a pair of impurities of the same chemical potential $\Gamma$.
We made the choice to introduce one impurity in B site and the other
in C site. This could symmetrize the wave function, thus allowing
resonance transport near energy $E=\Gamma/2$. Below we provide exact
wave function calculation for such impurity pair, and provide conditions
such that resonant can occur.

To begin with, we calculate the effect of a single impurity of strength
$\Gamma$, located on B site at $x=0$. \eqref{bx_a} is modified
as: 
\begin{equation}
	(E-\Gamma)b_{0}=J_{+}a_{0}+J_{-}a_{-1},
\end{equation}
where $a_{0}$ can be determined by \eqref{ax_2} by setting $x=0$
and $a_{-1}$ can be determined by \eqref{ax_1} by setting $x=-1$.
Additionally at C site for $x=-1$, using \eqref{cx_a} we have equation:
\begin{equation}
	Ec_{-1}=J_{+}a_{-1}+J_{-}a_{0}.
\end{equation}
By writing down the Schr\"odinger equation for A site at $x=0$ and
$x=-1$, keeping up to lowest order for $E\ll J\delta$, we additionally
have: 
\begin{align}
	J_{+}b_{-1}+J_{-}b_{0}+J_{-}c_{-2}+J_{+}c_{-1} & =0,\\
	J_{+}b_{0}+J_{-}b_{1}+J_{-}c_{-1}+J_{+}c_{0} & =0.
\end{align}
Solving all four equations gives us: 
\begin{align}
	b_{1} & =b_{-1}e^{4\delta},\\
	c_{0} & =c_{-2}e^{4\delta}+\frac{4\Gamma\delta}{2E-\Gamma}b_{-1}e^{2\delta}.
\end{align}
For an impurity on B site, on one hand, wavefunction of B orbital
is only affected exactly at the impurity position. On the other hand,
the wave function of C orbital has a discontinuous jump, between the
wavefunction on the left, and on the right of the impurity. This can
be regarded as a scattering event, in the Green's function language.
Similarly if we introduce a disorder at $x=n$ on orbital C, we have
a discontinuous jump at wave function of orbital B: 
\begin{equation}
	b_{n+2}=b_{n}e^{4\delta}-\frac{4\Gamma\delta}{2E-\Gamma}c_{n-1}e^{2\delta}.
\end{equation}
The wavefunction in between the pair of impurities can be related
by (\ref{bx}-\ref{cx}), using the decay factor $\kappa=(-1)^{n-1}e^{-2\delta(n-1)}$.
As such we obtain: 
\begin{equation}
	b_{n+2}=\kappa^{-1}e^{4\delta}b_{1}-\frac{4\delta\Gamma\kappa}{2E-\Gamma}c_{0}e^{2\delta}.
\end{equation}
Resonant occurs if $b_{n+2}=c_{-2}$ where the wavefunction is symmetrized.
We note that exactly at $E=\Gamma/2$, the wave function is asymmetrized
due to singularity, thus prohibiting transport. Define $\alpha=c_{-2}/b_{-1}$
we have two resonant peaks of energy: 
\begin{align}
	E_{\pm} & =\frac{\Gamma}{2}\pm\Delta E,\\
	\Delta E & =\Gamma\delta e^{-2\delta}\frac{\kappa^{2}\alpha}{\kappa\alpha-1}\left(\sqrt{1+\frac{4(1-\kappa\alpha)}{\kappa^{2}\alpha^{2}}e^{8\delta}}-1\right).
\end{align}
In the limit where $\alpha\kappa\ll1$, we get: 
\begin{equation}
	\Delta E\sim2\Gamma\delta e^{-2\delta(n-1)}
	\label{im_pair}
\end{equation}
Which is independent of the ration $\alpha$, agreeing with our argument
that resonant is due to symmetrizing of wave function. We note that
when disorder is introduced, the lengthscale interplays with the energy
scale. Phenomenologically, when we have random disorder of $\in[-\Gamma,\Gamma]$
to the whole lattice of length $L$, we introduce $L$ pairs of impurities,
each correspond to energy level $\Gamma_{i}$ with separation similar
to $\sim\Gamma/L$. In the weak disorder limit, the separation between
the energy levels, are smaller or similar to $\Delta E\propto\Gamma$,
thus there is strong interference, which is likely to be destructive
between different pairs of impurity. As such when we increase the
disorder strength, thus the separation, we weaken the destructive
interference and increase the maximal transmission. In the strong
disorder limit, the separation between the energy levels are much
larger than $\Delta E$. Thus when we average over the ensembles in
disorder calculation, it can be approximated as proportional to the
density of energy level $\sim1/\Gamma$. Additional discussion for
the strong disorder limit is included in the maintext.

\section{Green's function calculation}
\label{sec:S-Green}
To get more insights for the flatband transport, here we provide an alternative approach to calculating the transmission
profile, we show that the decay length of the bound state on a flat
band is determined by the band projector. Instead of a specific model, we consider a general local potential $V=\sum_{x\in\mathcal{V}}\sum_{\alpha\beta}V_{\alpha\beta}(x)c_{\alpha x}^{\dagger}c_{\beta x}$,
which acts on the local sites $\mathcal{V}$ in an infinite size 1D
lattice. Then the bound state wave function can be constructed from the
Lippmann-Schwinger equation with 
\begin{equation}
	\psi_{\alpha}(x)=\sum_{x^{\prime}\in\mathcal{V}}\sum_{\beta}G_{\alpha\beta}(x-x^{\prime},E)V_{\alpha\beta}(x^{\prime})\psi_{\beta}(x^{\prime}),
\end{equation}
where the $G_{\alpha\beta}(x,E)$ is the Green function of the free
part, 
\begin{equation}
	G_{\alpha\beta}(x,E)=\sum_{n}\int\frac{dk}{2\pi}e^{ikx}\frac{P_{n\alpha\beta}(k)}{E-\epsilon_{n0}(k)+i\zeta}
\end{equation}
with the band projector $\mathcal{P}_{n\alpha\beta}(k)=u_{n\beta}(k)u_{n\alpha}^{*}(k)$ under the band basis.
The long-distance behavior of the $\psi_{\alpha}(x)$ is controlled
by the asymptotic behavior of Green function $G(x,E)$ at large $x$.
As the flat band lacks dispersion, the decay length is exclusively
determined by the band projection. In one dimension, the band projection
has the tendency $e^{-h|x|}$ where $h$ is the distance of
a branch point from the real axis in the complex-$k$ plane.

Now we start to calculate the transmission
due to interface states using Green's function method. For a multiband system, in
the sublattice basis, it can be written as: 
\begin{equation}
	g_{\alpha,\beta}(\mathbf{r},\mathbf{r}';E)=\frac{1}{V_{\mathbf{k}}}\int d\mathbf{k}\ e^{i\mathbf{k}\cdot(\mathbf{r}-\mathbf{r}')}\sum_{i}\frac{[P_{i}(\mathbf{k})]_{\alpha\beta}}{E+i\zeta-\epsilon_{i}(\mathbf{k})},\label{eq:S-DefofGreenFun}
\end{equation}
where $\mathbf{r},\ \mathbf{r}'$ are the position vector of the lattice
site, $i,\ j$ are the band indices, $\alpha,\ \beta$ are the orbital
indices, $V_{\mathbf{k}}$ is the total volume of the first Brillouin
zone, $E$ is the energy, $\zeta\rightarrow0$ and $P_{i}(\mathbf{k})=\vert u_{i}(\mathbf{k})\rangle\langle u_{i}(\mathbf{k})\vert$
defines the projection matrix. For the diamond lattice, in particular
the flatband, the projection matrix is given by: 
\begin{equation}
	P_{f}(k)=\frac{1}{\epsilon_{0}^{2}(k)}\begin{pmatrix}0 & 0 & 0\\
		0 & \epsilon_{0}^{2}/2 & -(J_{+}e^{ik/2}+J_{-}e^{-ik/2})^{2}\\
		0 & -(J_{+}e^{-ik/2}+J_{-}e^{ik/2})^{2} & \epsilon_{0}^{2}/2
	\end{pmatrix}
\end{equation}
Similar to the wavefunction calculation, we will focus on $E\ll\Delta$
where $\Delta$ is the band gap. As such for the infinite Green's
function, only contribution from flat band is significant, which is
given by: 
\begin{equation}
	g_{\alpha,\beta}^{f}(n,n';E+i\zeta)=\frac{1}{2\pi}\int_{-\pi}^{\pi}dk\frac{e^{ik(n-n')}}{E+i\zeta}[P_{f}(k)]_{\alpha\beta}.
\end{equation}
We begin by studying the case where $\alpha=\beta\in\{B,C\}$: 
\begin{align}
	g_{\alpha,\alpha}^{f}(n,n';E+i\zeta) & =\frac{1}{2\pi}\int_{-\pi}^{\pi}dk\frac{e^{ik(n-n')}}{E+i\zeta}\frac{1}{2}\nonumber \\
	& =\frac{1}{2(E+i\zeta)}\left\{ \begin{array}{ll}
		1 & n=n'\\
		0 & n\neq n'
	\end{array}.\right.
\end{align}
As such for infinite size lattice, Green's function corresponding
to propagation between the same type of site due to flat band is always
$0$. We can also define the density of state for non-vanishing $\zeta$:
\begin{align}
	\rho_{\alpha}(E) & =-\frac{1}{\pi}\mathrm{Im}g_{\alpha,\alpha}^{f}(n,n;E+i\zeta)\nonumber \\
	& =\frac{\zeta}{2\pi(E^{2}+\zeta^{2})}.
\end{align}

Naively, if we interpret $\zeta$ as disorder, it has a band widening
effect on the flat band. Note that $\int_{-\infty}^{+\infty}dE\ \rho_{f}(E)=\frac{1}{2}$,
meaning electrons are evenly split between B and C sites. We begin
by deriving \eqref{g_BC}. The Green's function is defined by the
integral:

\begin{align}
	g_{BC}^{f}(n,n';E) & =-\frac{1}{2\pi}\int_{0}^{2\pi}dk\frac{e^{ik(n-n')}}{E+i\zeta}\frac{2J_{+}J_{-}+J_{+}^{2}e^{ik}+J_{-}^{2}e^{-ik}}{\epsilon(k)^{2}}\nonumber \\
	& =\int_{0}^{2\pi}dkf(k).\label{intf}
\end{align}
For convenience, we define $J_{+}=wJ$ and $J_{-}=w^{-1}J$ and change
the upper bound and lower bound to $[0,2\pi]$ according to $\int_{a}^{a+2\pi}f(k)dk=Const$
for periodic function $f(k)$. Note the pole $k_{\pm}$ of $f(k)$
are defined by $\epsilon(k_{\pm})=0$, which correspond to: 
\begin{equation}
	k_{\pm}=\pi\pm2i\ln w=\pi\pm i\lambda_{0}.
\end{equation}
Where we have defined $\lambda_{0}=2\ln w$. As such we can rewrite
the integral \eqref{intf} in terms of contour integral. For $n\geq n'$
we consider rectangular contour $C_{+}:0\rightarrow2\pi\rightarrow2\pi+i\infty\rightarrow i\infty\rightarrow0$:
\begin{equation}
	\int_{0}^{2\pi}dk\ f(k)=\oint dk\ f(k)-\int_{2\pi}^{2\pi+i\infty}dk\ f(k)-\int_{2\pi+i\infty}^{i\infty}dk\ f(k)-\int_{i\infty}^{0}dk\ f(k).
\end{equation}
Due to periodicity, we always have: 
\begin{equation}
	\int_{2\pi}^{2\pi+i\infty}dk\ f(k)=-\int_{i\infty}^{0}dk\ f(k).
\end{equation}
As such, two of the integrals cancel out with each other. The remaining
two integrals are given by: 
\begin{align}
	\int_{2\pi+i\infty}^{i\infty}dk\ f(k) & =-\frac{1}{2\pi}\lim_{\lambda\to+\infty}\int_{2\pi}^{0}dk\frac{e^{-\lambda(n-n')}e^{ik(n-n')}}{E+i\zeta}\frac{w^{2}e^{ik}e^{-\lambda}+w^{-2}e^{-ik}e^{\lambda}+2}{2(w^{2}+w^{-2}+2\cos(k+i\lambda))}\nonumber \\
	& =-\frac{1}{2\pi}\int_{2\pi}^{0}dk\frac{1}{E+i\zeta}\lim_{\lambda\rightarrow+\infty}\frac{e^{-\lambda(n-n')}e^{ik(n-n')}w^{-2}}{2}\nonumber \\
	& =\begin{cases}
		0 & \mathrm{if}\ n>n'\\
		\frac{1}{2w^{2}(E+i\zeta)} & \mathrm{if}\ n=n'
	\end{cases},\\
	\oint dk\ f(k) & =2\pi i\mathrm{Res}_{k\rightarrow\pi+i\lambda_{0}}f(k)\nonumber \\
	& =-i\mathrm{Res}_{k\rightarrow\pi+i\lambda_{0}}\frac{e^{ik(n-n')}}{E+i\zeta}\frac{2J_{+}J_{-}+J_{+}^{2}e^{ik}+J_{-}^{2}e^{-ik}}{2(J_{+}^{2}+J_{-}^{2}+2J_{+}J_{-}\cos k)}\nonumber \\
	& =-\frac{(-1)^{n-n'}e^{-\lambda_{0}(n-n')}}{2(E+i\zeta)}\mathrm{Res}_{\lambda=\lambda_{0}}\frac{2-w^{2}e^{-\lambda}-w^{-2}e^{\lambda}}{w^{2}+w^{-2}-2\cosh\lambda}\nonumber \\
	& =0.
\end{align}
As such we have $g_{BC}^{f}(n,n';E)=0$ for $n>n'$ and $g_{BC}^{f}(n,n;E)=-1/2w^{2}(E+i\zeta)$.
For $n<n'$, we use the contour $C_{-}:0\rightarrow2\pi\rightarrow2\pi-i\infty\rightarrow-i\infty\rightarrow0$
instead: 
\begin{equation}
	\int_{0}^{2\pi}dk\ f(k)=\oint dk\ f(k)-\int_{2\pi-i\infty}^{-i\infty}dk\ f(k).
\end{equation}
Note that: 
\begin{align}
	\int_{2\pi-i\infty}^{-i\infty}dk\ f(k) & =-\frac{1}{2\pi}\lim_{\lambda\to-\infty}\int_{2\pi}^{0}dk\frac{e^{-\lambda(n-n')}e^{ik(n-n')}}{E+i\zeta}\frac{w^{2}e^{ik}e^{-\lambda}+w^{-2}e^{-ik}e^{\lambda}+2}{2(w^{2}+w^{-2}+2\cos(k+i\lambda))}\nonumber \\
	& =-\frac{1}{2\pi}\int_{2\pi}^{0}dk\frac{1}{E+i\zeta}\lim_{\lambda\rightarrow\infty}\frac{e^{\lambda(n-n')}e^{ik(n-n')}w^{2}}{2}\nonumber \\
	& =0,\\
	\oint dk\ f(k) & =-2\pi i\mathrm{Res}_{k\rightarrow\pi-i\lambda_{0}}f(k)\nonumber \\
	& =i\mathrm{Res}_{k\rightarrow\pi+i\lambda_{0}}\frac{e^{ik(n-n')}}{E+i\zeta}\frac{2J_{+}J_{-}+J_{+}^{2}e^{ik}+J_{-}^{2}e^{-ik}}{2\left(J_{+}^{2}+J_{-}^{2}+2J_{+}J_{-}\cos k\right)}\nonumber \\
	& =\frac{(-1)^{n-n'}e^{\lambda_{0}(n-n')}}{2(E+i\zeta)}\mathrm{Res}_{\lambda=-\lambda_{0}}\frac{2-w^{4}-w^{-4}}{(w^{2}-w^{-2})(\lambda+\lambda_{0})}\nonumber \\
	& =-\frac{(-1)^{n-n'}w^{2(n-n')}}{2(E+i\zeta)}(w^{2}-w^{-2}).
\end{align}
To conclude for Green's function from B to C site, we have: 
\begin{equation}
	g_{BC}^{f}(n,n';E)=\begin{cases}
		(-1)^{n-n'}\frac{w^{2(n-n')}}{2(E+i\zeta)}(w^{2}-w^{-2}) & n<n'\\
		-\frac{1}{2(E+i\zeta)}\frac{1}{w^{2}} & n=n'\\
		0 & n>n'
	\end{cases}.\label{g_BC_2}
\end{equation}
By following the same procedure in deriving \eqref{g_BC_2} for C
to B site and substituting $w^{2}\sim(1+2\delta)$ for $\delta\ll1$
we can obtain:

\begin{align}
	g_{BC}^{f}(n,n';E) & \sim\begin{cases}
		-(-1)^{n-n'}2\frac{\delta}{E+i\zeta}e^{2\delta(n-n')} & n<n'\\
		-\frac{1}{2(E+i\zeta)} & n=n'\\
		0 & n>n'
	\end{cases},\label{g_BC}\\
	g_{CB}^{f}(n,n';E) & \sim\begin{cases}
		0 & n<n'\\
		-\frac{1}{2(E+i\zeta)} & n=n'\\
		-(-1)^{n-n'}2\frac{\delta}{E+i\zeta}e^{-2\delta(n-n')} & n>n'
	\end{cases}.\label{g_CB}
\end{align}

From the infinite Green's function between B and C site, we obtain
a decay length of $2\delta$, which is consistent with the decay length
obtained in wavefunction approach. Also we note that from B to C site
only forward direction gives non-zero result. This correspond to the
interface state localized on C site that decay in the forward direction.
Similarly from C to B site only backward direction gives non-zero
result. This correspond to the interface state localized on B site
that decay in the backward direction. Because of physicality, only
decay mode is allowed with an infinite lattice.

To calculate the semi-infinite Green's function for the diamond-like
lattice, we must consider the contribution from the dispersive band
on $A$ site, when $E\ll J\delta$. Recall the Hamiltonian is given
by 
\begin{equation}
	H(k)=\begin{pmatrix}0 & J_{+}+J_{-}e^{ik} & J_{+}+J_{-}e^{-ik}\\
		J_{+}+J_{-}e^{-ik} & 0 & 0\\
		J_{+}+J_{-}e^{ik} & 0 & 0
	\end{pmatrix}
\end{equation}
where the dispersive Bloch state is given by 
\begin{equation}
	\vket{u_{\pm}(k)}=\frac{1}{\sqrt{2}\epsilon(k)}\begin{pmatrix}\pm\epsilon(k)\\
		e^{-ik}J_{-}+J_{+}\\
		e^{ik}J_{-}+J_{+}
	\end{pmatrix}
\end{equation}
where the $\epsilon(k)=\sqrt{2\left(J_{+}^{2}+J_{-}^{2}+2J_{+}J_{-}\cos k\right)}$,
Thus the projector $P_{\pm,k}=\vket{u_{\pm}(k)}\vbra{u_{\pm}(k)}$
is given by 
\begin{equation}
	P_{\pm,k}=\frac{1}{4\epsilon(k)^{2}}\begin{pmatrix}2\epsilon(k)^{2} & \pm2\epsilon(k)(J_{-}e^{ik}+J_{+}) & \pm2\epsilon(k)(J_{-}e^{-ik}+J_{+})\\
		\pm2\epsilon(k)(J_{-}e^{-ik}+J_{+}) & \epsilon(k)^{2} & 2(J_{-}e^{-ik}+J_{+})^{2}\\
		\pm2\epsilon(k)(J_{-}e^{ik}+J_{+}) & 2(J_{-}e^{ik}+J_{+})^{2} & \epsilon(k)^{2}
	\end{pmatrix}
\end{equation}
Use eq(\ref{eq:S-DefofGreenFun}) we can calculate the Green's function
for A site on the same unit cell as 
\begin{equation}
	\begin{split}G_{AA}^{d}(n,n;E) & =\frac{1}{2\pi}\int_{-\pi}^{\pi}dk\frac{1}{2}\left(\frac{1}{E-\epsilon(k)+i\zeta}+\frac{1}{E+\epsilon(k)+i\zeta}\right)\\
		& =\frac{E+i\zeta}{2\pi}\int_{-\pi}^{\pi}dk\frac{1}{E^{2}-\epsilon(k)^{2}}\\
		& =-\frac{E+i\zeta}{\sqrt{(E^{2}-8J^{2})(E^{2}-8J^{2}\delta^{2})}}\\
		& \overset{E\ll J\delta}{\to}-\frac{E+i\zeta}{8J^{2}\delta}
	\end{split}
\end{equation}

\begin{equation}
	\begin{split}G_{AB}^{d}(n,n;E) & =\frac{1}{2\pi}\int_{-\pi}^{\pi}dk\frac{J_{-}e^{ik}+J_{+}}{2\epsilon(k)}\left(\frac{1}{E-\epsilon(k)+i\zeta}-\frac{1}{E+\epsilon(k)+i\zeta}\right)\\
		& =\frac{1}{2\pi}dk\int_{-\pi}^{\pi}\frac{J_{-}e^{ik}+J_{+}}{E^{2}-\epsilon(k)^{2}}\\
		& =-\frac{1}{2\pi}\frac{\pi\left(E^{2}+8J^{2}\delta+\sqrt{(E^{2}-8J^{2})(E^{2}-8J^{2}\delta^{2})}\right)}{2J\sqrt{(E^{2}-8J^{2})(E^{2}-8J^{2}\delta^{2})}(1+\delta)}\\
		& \overset{E\ll J\delta}{\to}-\frac{1}{2J(1+\delta)}\sim-\frac{1}{2J}
	\end{split}
\end{equation}
and the Green's function for A-C site is the same as A-B site 
\begin{equation}
	G_{AC}^{d}(n,n;E)=G_{AB}^{d}(n,n;E)=G_{BA}^{d}(n,n;E)=G_{CA}^{d}(n,n;E)
\end{equation}

The B site Green's function is given above and contribute mainly by
flat band near $E\sim0$ 
\begin{equation}
	\begin{split}G_{BB}^{d}(n,n;E) & =\frac{1}{2\pi}\int_{-\pi}^{\pi}dk\frac{1}{4}\left(\frac{1}{E-\epsilon(k)+i\zeta}+\frac{1}{E+\epsilon(k)+i\zeta}+\frac{2}{E+i\zeta}\right)\\
		& \overset{E\ll J\delta}{\to}-\frac{E+i\zeta}{8J^{2}\delta}+\frac{1}{2(E+i\zeta)}\sim\frac{1}{2(E+i\zeta)}
	\end{split}
\end{equation}

Now we need to calculate the Green's funcion of infinite diamond lattice.
We can cut the infinite diamond lattice into two semi-infinite segments.
We consider the left segment with a right boundary where the contribution
from C site is not important since it was isolated at the right boundary.
We can write down the Green's function for the right boundary with
only A and B site as

\begin{equation}
	\mathbf{G}_{0}=\begin{pmatrix}-\frac{E+i\zeta}{8J^{2}\delta} & -\frac{1}{2J}\\
		-\frac{1}{2J} & \frac{1}{2E+i\zeta}
	\end{pmatrix}
\end{equation}

Now, follow the Dyson's equation, we can write down the Green's function
with another sublattice attached to it as 
Since we are considering an semi infinite chain, we should expect
the new Green's function should be the same as the previous one. Notice
that the hopping matrix with two blocks are 
\begin{equation}
	\mathbf{V}=\begin{pmatrix}0 & J_{-}\\
		J_{-} & 0
	\end{pmatrix}
\end{equation}
Use the Dyson equation, we can get the total Green's function $\mathbf{G}$
as 
\begin{equation}
	\mathbf{G}=\mathbf{G}_{0}+\mathbf{G}_{0}\mathbf{\Sigma}\mathbf{G}
\end{equation}
For the boundary lead coupling, the self-energy has the form 
\begin{equation}
	\mathbf{\Sigma}=V\mathbf{G}V^{\dagger}
\end{equation}
plug this into the Dyson equation and solve the equation to the lowest
order of $\delta$ and $E$ we can get 
\begin{equation}
	\mathbf{G}\sim\begin{pmatrix}\frac{E}{J^{2}} & \frac{i}{J}\\
		\frac{i}{J} & -\frac{4\delta}{E}
	\end{pmatrix}
\end{equation}

Then we consider to connect the Leads to the diamond lattice. For simplicity
we assume bothe the left and right lead are 1D chain with nearest
neighbor hopping $t_{N}$. The semi-infinite Green's function can
be calculated as 
\begin{equation}
	\begin{split}g_{L} & =\frac{1}{2t_{N}^{2}}\left(E-i\sqrt{4t_{N}^{2}-E^{2}}\right)\\
		& \overset{E\ll t_{N}}{\sim}\frac{E}{2t_{N}^{2}}-\frac{i}{t_{N}}
	\end{split}
	\label{eq:SGF_Lead}
\end{equation}

Where we have assumed $E\ll t_{N}$. For the diamond-like lattice, with
finite size $L$, the Green's function can be approximated as: 
\begin{align}
	\bar{g}(E) & \sim\mathbf{G}_{BB}\sim\frac{4\delta}{E},\label{gbar}\\
	|g_{p}(E)| & \sim\frac{8\delta}{E}e^{-2\delta L}.
\end{align}
Where $\bar{g}$ is the same site Green's function for the sublattice
that is coupled to the lead, and $g_{p}$ is the Green's function
correspond to transport between the two end of the lattice. Note that
we have assumed that $E\gg\zeta$, namely the effect of disorder is
negligible. When coupled to the lead, the new propagation Green's
function $G(E)$ is given by Dyson equation again:

\begin{equation}
	G(E)=g_{p}+g_{p}\mathbf{\Sigma}G(E),
\end{equation}
where the self energy is given by 
\begin{equation}
	\mathbf{\Sigma}=V\bar{g}g_{L}V^{\dagger}=T_{\partial}^{2}\bar{g}g_{L}.
\end{equation}
Given that the diamond-like lattice is long enough (i.e. $\delta L>1$,
where $L$ is the number of site), $g_{p}\ll\bar{g}$ and we assume
finite size only change the prefactor of $g_{p}$, and the derivation
for $\bar{g}$ is given by \eqref{gbar}, then we can write down the
dressed Green's function which we represent diagrammatically as the
following diagram up to the first order of $g_{p}$: 
\begin{figure}[H]
	\centering \includegraphics[width=0.7\linewidth]{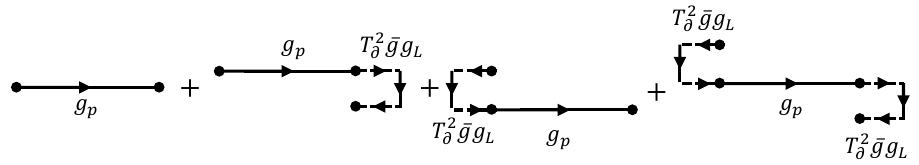}. 
\end{figure}

The diagrams lead to 
\begin{equation}
	\begin{split}G(E) & =g_{p}+g_{p}\mathbf{\Sigma}^{\dagger}+\mathbf{\Sigma}g_{p}+\mathbf{\Sigma}g_{p}\mathbf{\Sigma}^{\dagger}+O(g_{p}^{3})\\
		& =g_{p}+g_{p}\left[2T_{\partial}^{2}\bar{g}g_{L}+(T_{\partial}^{2}\bar{g}g_{L})^{2}\right]+\cdots\\
		& \sim\frac{g_{p}}{\mathcal{D}},
	\end{split}
\end{equation}
where $\mathcal{D}$ is the Dyson factor and $T_{\partial}$ is the
contact hopping strength 
\begin{align}
	\mathcal{D} & \approx1-2T_{\partial}^{2}\bar{g}g_{L}+3(T_{\partial}^{2}\bar{g}g_{L})^{2}.
\end{align}
Since the contact hopping strength is small, we can approximate the
prefactor as 
\begin{align}
	\mathcal{D} & \sim\left(T_{\partial}^{2}g_{L}\bar{g}-1\right)^{2}\nonumber \\
	& \sim\left(1+\frac{4T_{\partial}^{2}\delta}{t_{N}}\frac{i}{E}\right)^{2},\\
	|G(E)|^{2} & \sim\frac{64\delta^{2}e^{-4\delta L}}{(E^{2}+E_{0}^{2})^{2}}E^{2}.
\end{align}

Where $E_{0}=4T_{\partial}^{2}\delta/t_{N}$, which is same
as the $E_{0}$ we have defined in the wavefunction calculation. We
can determine the transmittance in terms of the Green's function according
to Fisher-Lee relation: 
\begin{align}
	\mathcal{T} & =\mathrm{Tr}(\Gamma_{L}G\Gamma_{R}G^{\dagger})\nonumber \\
	& \sim|G|^{2}\left(\frac{2T_{\partial}^{2}}{t_{N}}\right)^{2}\nonumber \\
	& \sim \frac{16E^2E_0^2 e^{-4\delta L}}{(E^2+E_0^2)^2}
	.
\end{align}
where the spread function $\Gamma=i(\Sigma-\Sigma^{\dagger})=iT_{\partial}^{2}(g_{L}-g_{L}^{\dagger})$
is defined by the Green's function of the lead in eq(\ref{eq:SGF_Lead}).
We can see that the transmittance is identical to the result from
wavefunction calculation \eqref{eq:transmittance} in the long junction
limit.


\section{Kubo-Greenwood formula and conductivity in the clean limit}
\subsection{Derivation on Kubo-Greenwood formula}
The Kubo-formula gives the conductivity as \cite{S-Book-Bruus}
\begin{equation}
	\sigma^{\alpha\beta}(\mathbf{r},\mathbf{r}';\omega)=\frac{ie^{2}}{\omega}\Pi_{\alpha\beta}^{R}(\mathbf{r},\mathbf{r}';\omega)+\frac{ie^{2}n(\mathbf{r})}{\omega m}\delta(\mathbf{r}-\mathbf{r}')\delta_{\alpha\beta},
\end{equation}
where $n(\mathbf{r})$ is the particle density and $\Pi$ is the
current-current correlation function 
\begin{equation}
	\Pi_{\alpha\beta}^{R}(\mathbf{r},\mathbf{r}';t-t')=C_{J^{\alpha}(\mathbf{r})J^{\beta}(\mathbf{r'})}^{R}(t-t')=-i\theta(t-t')\average{\left[J^{\alpha}(\mathbf{r},t),J^{\beta}(\mathbf{r}',t')\right]}_{0}.
\end{equation}

Use the many-body eigenstate $H\vket{n}=E_{n}\vket{n}$, we can write
down the fourier transform of current-current correlation function
as 
\begin{equation}
	\begin{split}\Pi_{\alpha\beta}^{R} & (\mathbf{r},\mathbf{r}';\omega)=-i\int dt\,e^{i\omega t}\theta(t)\average{\left[J^{\alpha}(\mathbf{r},t),J^{\beta}(\mathbf{r}',0)\right]}_{0}\\
		& =-i\frac{1}{\mathcal{Z}_{0}}\sum_{n}\int_{0}^{\infty}dt\,e^{i\omega t}\vbra{n}\left[e^{iH_{0}t}J^{\alpha}(\mathbf{r})e^{-iH_{0}t},J^{\beta}(\mathbf{r}')\right]\vket{n}e^{-\beta E_{n}}\\
		& =-i\frac{1}{\mathcal{Z}_{0}}\sum_{mn}\int_{0}^{\infty}dt\,e^{i\omega t}\left(e^{i(E_{n}-E_{m})t}\vbra{n}J^{\alpha}(\mathbf{r})\vket{m}\vbra{m}J^{\beta}(\mathbf{r}')\vket{n}-e^{-i(E_{n}-E_{m})t}\vbra{m}J^{\alpha}(\mathbf{r})\vket{n}\vbra{n}J^{\beta}(\mathbf{r}')\vket{m}\right)e^{-\beta E_{n}}\\
		& =-i\frac{1}{\mathcal{Z}_{0}}\sum_{mn}\int_{0}^{\infty}dt\,e^{i(\omega+E_{n}-E_{m})t}\left(e^{-\beta E_{n}}-e^{-\beta E_{m}}\right)\vbra{n}J^{\alpha}(\mathbf{r})\vket{m}\vbra{m}J^{\beta}(\mathbf{r}')\vket{n}.
	\end{split}
	\label{eq:ParaFourier}
\end{equation}

For most materials we can assume the electrons are non-interacting,
where the many-body hamiltonian can be reduced to the sum of single
body Hamiltonian $\sum_{i}H_{0}(i)$ so we can use the single body
eigenstate $H_{0}\vket{n}=\epsilon_{n}\vket{n}$ and the current operator
can be written as\cite{S-Book-allen}

\begin{equation}
	J^{\alpha}(\mathbf{r})=\sum_{\mu\nu}\vbra{\mu}J_{\alpha}^{(1)}(\mathbf{r})\vket{\nu}a_{\mu}^{\dagger}a_{\nu}.
\end{equation}

we plug it into eq(\ref{eq:ParaFourier}), we focus on the term $e^{-\beta E_{n}}$
\begin{equation}
	\begin{split} & \sum_{mn}\frac{e^{-\beta E_{n}}}{\mathcal{Z}_{0}}e^{i(E_{n}-E_{m})t}\vbra{n}J^{\alpha}(\mathbf{r})\vket{m}\vbra{m}J^{\beta}(\mathbf{r}')\vket{n}\\
		& =\sum_{mn,\mu\nu\rho\sigma}\frac{e^{-\beta E_{n}}}{\mathcal{Z}_{0}}e^{i(E_{n}-E_{m})t}\vbra{\mu}J_{\alpha}^{(1)}(\mathbf{r})\vket{\nu}\vbra{\rho}J_{\beta}^{(1)}(\mathbf{r}')\vket{\sigma}\vbra{n}a_{\mu}^{\dagger}a_{\nu}\vket{m}\vbra{m}a_{\rho}^{\dagger}a_{\sigma}\vket{n}
	\end{split}
\end{equation}

Notice that the expectation 
\[
\vbra{n}a_{\mu}^{\dagger}a_{\nu}\vket{m}\vbra{m}a_{\rho}^{\dagger}a_{\sigma}\vket{n}
\]
are non-zero only when $\mu=\sigma,\nu=\rho$ or $\mu=\nu,\rho=\sigma$.
Which lead to $E_{n}-E_{m}=\epsilon_{\mu}-\epsilon_{\nu}$ or $E_{n}-E_{m}=0$.
Now we get 
\begin{equation}
	\begin{split} & \sum_{n,\mu\nu\rho\sigma}\frac{e^{-\beta E_{n}}}{\mathcal{Z}_{0}}e^{i(\epsilon_{\mu}-\epsilon_{\nu})t}\vbra{\mu}J_{\alpha}^{(1)}(\mathbf{r})\vket{\nu}\vbra{\rho}J_{\beta}^{(1)}(\mathbf{r}')\vket{\sigma}\vbra{n}a_{\mu}^{\dagger}a_{\nu}\left(\sum_{m}\vket{m}\vbra{m}\right)a_{\rho}^{\dagger}a_{\sigma}\vket{n}\\
		& =\sum_{\mu\nu\rho\sigma}e^{i(\epsilon_{\mu}-\epsilon_{\nu})t}\vbra{\mu}J_{\alpha}^{(1)}(\mathbf{r})\vket{\nu}\vbra{\rho}J_{\beta}^{(1)}(\mathbf{r}')\vket{\sigma}\left(\sum_{n}\frac{e^{-\beta E_{n}}}{\mathcal{Z}_{0}}\vbra{n}a_{\mu}^{\dagger}a_{\nu}a_{\rho}^{\dagger}a_{\sigma}\vket{n}\right)\\
		& =\sum_{\mu\nu\rho\sigma}e^{i(\epsilon_{\mu}-\epsilon_{\nu})t}\vbra{\mu}J_{\alpha}^{(1)}(\mathbf{r})\vket{\nu}\vbra{\rho}J_{\beta}^{(1)}(\mathbf{r}')\vket{\sigma}\average{a_{\mu}^{\dagger}a_{\nu}a_{\rho}^{\dagger}a_{\sigma}},
	\end{split}
\end{equation}
we still need to calculate the expectation of ladder operators as
\begin{equation}
	\begin{split}\average{a_{\mu}^{\dagger}a_{\nu}a_{\rho}^{\dagger}a_{\sigma}} & =\average{a_{\mu}^{\dagger}a_{\nu}}\average{a_{\rho}^{\dagger}a_{\sigma}}+\average{a_{\mu}^{\dagger}a_{\sigma}}\average{a_{\nu}a_{\rho}^{\dagger}}\\
		& =f_{\mu}f_{\rho}\delta_{\mu\nu}\delta_{\rho\sigma}+f_{\mu}(1-f_{\nu})\delta_{\mu\sigma}\delta_{\nu\rho},
	\end{split}
\end{equation}
where $f_{\mu}=f(\epsilon_{\mu})$ is Fermi distribution function.
\begin{equation}
	\begin{split}\sum_{\mu\rho}\left[\vbra{\mu}J_{\alpha}^{(1)}(\mathbf{r})\vket{\mu}\vbra{\rho}J_{\beta}^{(1)}(\mathbf{r}')\vket{\rho}f_{\mu}f_{\rho}+e^{i(\epsilon_{\mu}-\epsilon_{\rho})t}\vbra{\mu}J_{\alpha}^{(1)}(\mathbf{r})\vket{\rho}\vbra{\rho}J_{\beta}^{(1)}(\mathbf{r}')\vket{\mu}f_{\mu}(1-f_{\rho})\right]\end{split}
\end{equation}

Similarly, we can obtain the result for term $e^{-\beta E_{m}}$ as
\begin{equation}
	\begin{split}\sum_{\mu\rho}\left[\vbra{\mu}J_{\alpha}^{(1)}(\mathbf{r})\vket{\mu}\vbra{\rho}J_{\beta}^{(1)}(\mathbf{r}')\vket{\rho}f_{\mu}f_{\rho}+e^{i(\epsilon_{\mu}-\epsilon_{\rho})t}\vbra{\mu}J_{\alpha}^{(1)}(\mathbf{r})\vket{\rho}\vbra{\rho}J_{\beta}^{(1)}(\mathbf{r}')\vket{\mu}f_{\rho}(1-f_{\mu})\right].\end{split}
\end{equation}

Collect these result together and we can get the conductivity in single
particle basis as 
\begin{equation}
	\sigma^{\alpha\beta}(\mathbf{r},\mathbf{r}';\omega)=\frac{ie^{2}n(\mathbf{r})}{\omega m}\delta(\mathbf{r}-\mathbf{r}')\delta_{\alpha\beta}+\frac{ie^{2}}{\omega}\sum_{\mu\rho}(f_{\mu}-f_{\rho})\frac{\vbra{\mu}J_{\alpha}^{(1)}(\mathbf{r})\vket{\rho}\vbra{\rho}J_{\beta}^{(1)}(\mathbf{r}')\vket{\mu}}{\omega+\epsilon_{\mu}-\epsilon_{\rho}+i\eta}
\end{equation}

Now the second term can be separated by using the expansion $\frac{1}{\omega(\omega+\Delta)}=\frac{1}{\Delta}\left(\frac{1}{\omega}-\frac{1}{\omega+\Delta}\right)$,
with the definition of single body current operator 
\begin{equation}
	\hat{\mathbf{J}}_{i}^{(1)}(\mathbf{r})=\frac{1}{2m}\left[\hat{\mathbf{p}}_{i}\delta(\hat{\mathbf{r}}_{i}-\mathbf{r})+\delta(\hat{\mathbf{r}}_{i}-\mathbf{r})\hat{\mathbf{p}}_{i}\right].\label{eq:CurrentOperator}
\end{equation}
we can show that in momentum space the current operator reduced to
momentum operator in the uniform limit 
\begin{equation}
	\hat{\mathbf{J}}(\mathbf{q}=0)=\frac{\hat{\mathbf{p}}}{m}.
\end{equation}
With the f-sum rule in momentum space 
\begin{equation}
	\sum_{\mu\rho}\frac{f_{\mu}-f_{\rho}}{\epsilon_{\mu}-\epsilon_{\rho}}\vbra{\mu}p_{\alpha}\vket{\rho}\vbra{\rho}p_{\beta}\vket{\mu}=-mn\delta_{\alpha\beta},
\end{equation}
after some algebra, it can be shown that the diamagnetic term was
cancelled and we can arrive at the Kubo-Greenwood formula under uniform limit $\mathbf{q}=0$
\begin{equation}
	\label{eq:S-Kubo-Greenwood}
	\begin{split}\sigma^{\alpha\beta}(\mathbf{0};\omega)=\frac{-ie^{2}}{m^{2}V}\sum_{\mu\rho}\frac{f_{\mu}-f_{\rho}}{\epsilon_{\mu}-\epsilon_{\rho}}\frac{\vbra{\mu}p_{\alpha}\vket{\rho}\vbra{\rho}p_{\beta}\vket{\mu}}{\omega+\epsilon_{\mu}-\epsilon_{\rho}+i\eta}.\end{split}
\end{equation}

Now it's safe to take DC limit and use the completeness relation

\begin{equation}
	\begin{split}\sigma^{\alpha\beta}(\mathbf{0};0) & =\frac{-ie^{2}}{m^{2}V}\sum_{\mu\rho}\frac{f_{\mu}-f_{\rho}}{\epsilon_{\mu}-\epsilon_{\rho}}\frac{\vbra{\mu}p_{\alpha}\vket{\rho}\vbra{\rho}p_{\beta}\vket{\mu}}{\epsilon_{\mu}-\epsilon_{\rho}+i\eta}\\
		& =\frac{e^{2}\pi}{m^{2}V}\int d\mathbf{r}d\mathbf{r}'\,\sum_{\mu\rho}\left(-\pd{f(\epsilon_{\mu})}{\epsilon_{\mu}}\right)\delta(\epsilon_{\mu}-\epsilon_{\rho})\left[\psi_{\mu}^{*}(\mathbf{r})p_{\mathbf{r}}^{\alpha}\psi_{\rho}(\mathbf{r})\right]\left[\psi_{\rho}^{*}(\mathbf{r}')p_{\mathbf{r}'}^{\beta}\psi_{\mu}(\mathbf{r}')\right]
	\end{split}
\end{equation}
and insert the identity $1=\int dE\delta(E-\epsilon_{\mu})$ 
\begin{equation}
	\frac{e^{2}\pi}{m^{2}}\int dE\left(-\pd{f(E)}{E}\right)\sum_{\mu\rho}\delta(E-\epsilon_{\rho})\delta(E-\epsilon_{\mu})\left[\psi_{\mu}^{*}(\mathbf{r})p_{\mathbf{r}}^{\alpha}\psi_{\rho}(\mathbf{r})\right]\left[\psi_{\rho}^{*}(\mathbf{r}')p_{\mathbf{r}'}^{\beta}\psi_{\mu}(\mathbf{r}')\right]
\end{equation}
it can be show that for single particle Green's function there's an
identity 
\begin{equation}
	-\frac{1}{\pi}\Im G(\mathbf{r},\mathbf{r}';E)=\sum_{\mu}\psi_{\mu}(\mathbf{r})\psi_{\mu}^{*}(\mathbf{r}')\delta(E-\epsilon_{\mu})
\end{equation}
use the cyclic symmetry of trace and we can arrive at the Kubo-Greenwood
formula 
\begin{equation}
	\sigma^{\alpha\beta}=\frac{e^{2}}{\pi V}\int dE\left(-\frac{\partial f}{\partial E}\right)\trace\left[\Im G(E)\hat{v}^{\alpha}\Im G(E)\hat{v}^{\beta}\right]\label{eq:Kubo-Streda}
\end{equation}
where we have defined the velocity operator $\hat{v}^{\alpha}=\frac{\hat{p}^{\alpha}}{m}$ to absorb the mass factor and the integral over $d\mathbf{r}$ and $d\mathbf{r}'$ is included in the trace.

\subsection{Transport in the clean limit}
We first apply the Kubu-Greenwood formula Eq. (\ref{eq:S-Kubo-Greenwood}) in the clean limit\cite{Kubo-Greenwood,S-Book-Meso}. For convenience, we do it under the band basis, where the Kubo-Greenwood formula becomes
\begin{equation}
	\sigma^{\alpha\beta}(\omega)=\frac{-ie^{2}}{V}\sum_{\mathbf{k}}\sum_{mn}\frac{f(\epsilon_{m}(\mathbf{k}))-f(\epsilon_{n}(\mathbf{k}))}{\epsilon_{m}(\mathbf{k})-\epsilon_{n}(\mathbf{k})}\frac{[v^{\alpha}(\mathbf{k})]_{mn}[v^{\beta}(\mathbf{k})]_{nm}}{\omega+\epsilon_{m}(\mathbf{k})-\epsilon_{n}(\mathbf{k})+i\eta},
\end{equation}
where a positive infinitesimal $\eta$ is introduced for small scattering rate. Since we are discussing the 1D chain, we only need to consider $ \alpha=\beta=x$ so that we can omit the direction index. And the components of velocity operator becomes
\begin{equation}
	\label{eq:S-velocityOperator}
	\begin{split}
		[v(k)]_{mn}&=\vbra{u_{mk}}i[\hat{H}_{k},\hat{r}]\vket{u_{nk}}=\vbra{u_{mk}}\nabla_{k}\hat{H}_{k}\vket{u_{nk}}\\
		&=\delta_{mn}\pd{\epsilon_{n}(k)}{k}-(\epsilon_{m}(k)-\epsilon_{n}(k))\vbra{u_{mk}}\ket{\partial_{k}u_{nk}}.
	\end{split}
\end{equation}

We first discuss the intraband contribution, where the conductivity becomes 
\begin{equation}
	\sigma^{\text{intra}}(\omega)=\frac{-ie^{2}}{ L}\sum_{k}\sum_{n}\left.\pd{f}{\epsilon}\right|_{\epsilon=\epsilon_{n}(k)}\frac{[v(k)]_{nn}[v(k)]_{nn}}{\omega+i\eta}.
\end{equation}
It can be shown that for a flatband, the intraband velocity operator will vanish
\begin{equation}
	[v(k)]_{nn}=\pd{\epsilon_{\text{flat}}}{k}=0,
\end{equation} 
which means the intraband contribution is zero in a flat band when $T=0$. If we tune the chemical potential to be in the flat band, the intraband contribution from other dispersive band will also be 0 since the derivative of Fermi distrubution only picks up the contrubution from flat band. So we can say that the intraband contribution is zero in the clean limit.

Then we need to calculate the interband contribution, that is
\begin{equation}
	\begin{split}
		\sigma^{\text{inter}}(\omega)&=\frac{-ie^{2}}{ L}\sum_{k}\sum_{m\neq n}\frac{f(\epsilon_{m}(k))-f(\epsilon_{n}(k))}{\epsilon_{m}(k)-\epsilon_{n}(k)}\frac{[v(k)]_{mn}[v(k)]_{nm}}{\omega+\epsilon_{m}(k)-\epsilon_{n}(k)+i\eta}\\
		&=\frac{-ie^{2}}{ L}\sum_{k}\sum_{m\neq n}\frac{f(\epsilon_{m}(k))-f(\epsilon_{n}(k))}{\omega+\epsilon_{m}(k)-\epsilon_{n}(k)+i\eta}\vbra{\partial_{k}u_{mk}}\ket{u_{nk}}\vbra{u_{nk}}\ket{\partial_{k}u_{mk}}(\epsilon_{m}(k)-\epsilon_{n}(k)).
	\end{split}
\end{equation}
We focus on the real part of the conductivity and make use of $\partial_{k}(\vbra{u_{mk}}\ket{u_{nk}})=0$ to get
\begin{equation}
	\begin{split}
		\Re	\sigma^{\text{inter}}(\omega)&=\frac{-e^{2}}{ L}\sum_{k}\sum_{m\neq n}f(\epsilon_{m})(\epsilon_{m}-\epsilon_{n})\vbra{\partial_{k}u_{mk}}\ket{u_{nk}}\vbra{u_{nk}}\ket{\partial_{k}u_{mk}}\left[\frac{\eta}{\left(\omega+\epsilon_{m}-\epsilon_{n}\right)^{2}+\eta^{2}}+
		\frac{\eta}{\left(\omega+\epsilon_{n}-\epsilon_{m}\right)^{2}+\eta^{2}}\right],
	\end{split}
\end{equation}

For the real part, we take the clean limit $\eta\to 0$ and make use of the limit$\lim\limits_{\eta\to0^{+}}\frac{\eta}{x^{2}+\eta^{2}}=\pi\delta(x)$ to rewrite the real part as
\begin{equation}
	\Re	\sigma^{\text{inter}}(\omega)=\frac{-e^{2}{\pi}}{ L}\sum_{k}\sum_{m\neq n}f(\epsilon_{m})(\epsilon_{m}-\epsilon_{n})\vbra{\partial_{k}u_{mk}}\ket{u_{nk}}\vbra{u_{nk}}\ket{\partial_{k}u_{mk}}\left[\delta(\omega-(\epsilon_{n}-\epsilon_{m}))+\delta(\omega-(\epsilon_{m}-\epsilon_{n}))\right],
\end{equation}
If all bands are isolated from each other, the conductivity will be 0 in the DC limit. So there's no DC transport in the clean limit according to the Kubo-Greenwood formula.

For the imaginary part, we can write it as 
\begin{equation}
	\Im	\sigma^{\text{inter}}(\omega)=\frac{-e^{2}}{ L}\sum_{k}\sum_{m\neq n}f(\epsilon_{m})(\epsilon_{m}-\epsilon_{n})\vbra{\partial_{k}u_{mk}}\ket{u_{nk}}\vbra{u_{nk}}\ket{\partial_{k}u_{mk}}
	\left[
	\frac{(\omega+\epsilon_{m}-\epsilon_{n})}{\left(\omega+\epsilon_{m}-\epsilon_{n}\right)^{2}+\eta^{2}}
	+\frac{(\omega+\epsilon_{n}-\epsilon_{m})}{\left(\omega+\epsilon_{n}-\epsilon_{m}\right)^{2}+\eta^{2}}
	\right].
\end{equation}
it's safe to take $\eta\to0$ to reduce the result
\begin{equation}
	\Im	\sigma^{\text{inter}}(\omega)=\frac{-2e^{2}}{ L}\sum_{k}\sum_{m\neq n}f(\epsilon_{m})\frac{\omega(\epsilon_{m}-\epsilon_{n})}{\omega^{2}-(\epsilon_{m}-\epsilon_{n})^{2}}\vbra{\partial_{k}u_{mk}}\ket{u_{nk}}\vbra{u_{nk}}\ket{\partial_{k}u_{mk}},
\end{equation}
and in the DC limit the result vanishes.

\section{Diagrammatic calculation on the disorder}

\subsection{Model Hamiltonian}

We need to calculate the transport in flat band system\cite{S-Book-Coleman}. We start from
the 1D Hamiltonian,
\begin{equation}
	\begin{split}H & =H_{0}+V\\
		& =\sum_{\alpha\beta}\int dxdx'\,\psi_{\alpha}^{\dagger}(x)H_{\alpha\beta}^{0}(x-x')\psi_{\beta}(x')+\sum_{\alpha}\int dx\,U(x)\psi_{\alpha}^{\dagger}(x)\psi_{\alpha}(x)
	\end{split}
\end{equation}
with $\psi$ being the Fermionic operator and $\alpha$ as orbital
degrees of freedom. The $U(x)$ represents the real scattering	potential generated by $N$ impurities distributed randomly, 
\begin{equation}
	U(x)=\sum_{j}\mathcal{U}(x-X_{j}).\label{eq:S-disorder_sum}
\end{equation}
We can perform the Fourier transform $\psi_{\alpha}(x)=\frac{1}{\sqrt{N}}\sum_{k}e^{ikx}c_{k,\alpha}$
and the Hamiltonian becomes 
\begin{equation}
	H=\sum_{k}\sum_{\alpha\beta}h_{\alpha\beta}(k)c_{k,\alpha}^{\dagger}c_{k,\beta}+\frac{1}{N}\sum_{kq}\sum_{\alpha}U(q)c_{k,\alpha}^{\dagger}c_{k-q,\alpha},
\end{equation}
where we defined the Fourier transform of the kernel of Hamiltonian
and disordered potential as 
\begin{equation}
	H^0_{\alpha\beta}(x-x')=\frac{1}{N}\sum_{p}e^{ip(x-x')}h_{\alpha\beta}(p),\quad U(x)=\frac{1}{N}\sum_{k}\,U(k)e^{ikx}
\end{equation}
The free Hamiltonian can be diagonalized and give out the bands 
\begin{equation}
	h_{\alpha\beta}(k)u_{\beta}^n(k)=\epsilon_n(k) u_{\alpha}^n(k),
\end{equation}
where $u_{\alpha}^n(k)$ is the $\alpha$ component of Bloch eigenstate $\vket{u_n(k)}$ of the free Hamiltonian $h_k$ and $\epsilon_n(k)$ is the energy bands at the corresponding momentum $k$. With the energy band basis, we build up a unitary matrix $U_{\alpha m}=u^m_\alpha$ for each momentum $k$ to diagonalize the free Hamiltonian and the total Hamiltonian can be writtern as
\begin{equation}\label{eq:S-Hamiltonian-band}
	H=\sum_{kn}\epsilon_{n}(k)c_{k,n}^{\dagger}c_{k,n}+\sum_{kq}\sum_{mn}\Gamma_{mn}(k,q)c_{k,m}^{\dagger}c_{k-q,n},\quad c_{k,n}=\sum_{\alpha}U^*_{n\alpha}(k)c_{k,\alpha},
\end{equation}
with the form factor $\Gamma_{mn}(k,q)=\frac{U(q)}{N}\vbra{u_{m}(k)}\ket{u_{n}(k-q)}$.
To proceed, we need to clarify the disorder potential. In this section,
we refer to the disorder average as
\begin{equation}
	\overline{\mathcal{O}}_{dis}=\int\prod_{j}\frac{1}{N}dX_{j}\mathcal{O}(X_{j})
\end{equation}
Using Eq.~(\ref{eq:S-disorder_sum}), we can write the potential as 
\begin{equation}
	U(k)
	=\sum_{j}e^{-ikX_{j}}\int dx\,\mathcal{U}(x-X_{j})e^{-ik(x-X_{j})}
	=\mathcal{U}(k)\sum_{j}e^{-ikX_{j}}.
\end{equation}
For simplicity, we assume the chemical potential shift due to disorder
potential is zero, i.e. $\average{U(x)}_{dis}=0$. In this
situation, the fluctuations are 
\begin{align}
	\overline{U(x)U(x')}_{dis} & =\frac{1}{N^{2}}\sum_{kk'}e^{ikx+ik'x'}\overline{U(k)U(k')}_{dis}\nonumber \\
	&  =\frac{1}{N^{2}}\sum_{kk'}\sum_{lm}e^{ikx+ik'x'}\mathcal{U}(k)\mathcal{U}(k')\overline{(e^{-ikX_{l}-ik'X_{m}})}_{dis}
\end{align}
Since the phase terms are independent at different sites, the disorder
average is nonzero only when $l=m$. Then we have 
\begin{equation}
	\sum_{lm}\overline{(e^{-ikX_{l}-ik'X_{m}})}_{dis}=\frac{N_{imp}}{N}\delta_{k,-k'}=n_{imp}\delta_{k,-k'}
\end{equation}
So we have
\begin{align}
	\overline{U(x)U(x')}_{dis} & =\frac{n_{imp}}{N^{2}}\sum_{k}e^{ik(x-x')}\abs{\mathcal{U}(k)}^{2}\\
	\overline{U(k)U(k')}_{dis} & =n_{imp}\abs{\mathcal{U}(k)}^{2}\delta_{k,-k'}
\end{align}
If we assume the impurity scattering is dominated by low-energy scattering, the scattering potential $\abs{\mathcal{U}(k)}^2\sim \abs{\mathcal{U}}^2$ and define $\gamma^2=\frac{n_{imp}}{N}\abs{\mathcal{U}}^2$,
we can approximate the fluctuation as 
\begin{align}
	\overline{U(x)U(x')}_{dis}&\sim \gamma^{2}\delta(x-x')\\
	\overline{U(k)U(k')}_{dis}&\sim \gamma^{2}N\delta_{k,-k'}.
\end{align}
where we have absorbed the $n_{imp}$ into $\gamma^2$ factor. 
For simplicity, we omit the subscript '$dis$' for the disorder average $\overline{\mathcal{O}}_{dis}$ and denote it by simply $\overline{\mathcal{O}}$ in the remaining part.

\subsection{Disorder-averaged Green's function}

In our setup, we assume the disorder strength is much weaker than
the band gap. We assume that conventional diagrammatic techniques
may be applied. For the single-particle Green function, we apply the
self-consistent Born approximation to the Green function. With a large
band gap which separates the flat band with others, we can ignore the
other bands's correction to the flatband Green function. Hence, we
can formulate the Dyson's equation for $\overline{G_{00}(k,\omega)}$,
\begin{equation}
	\overline{G_{00}(k,\omega)}=G_{00}^{0}(k,\omega)+ G_{00}^{0}(k,\omega)\Sigma(k,\omega)\overline{G_{00}(k,\omega)},
\end{equation}
with the self-energy 
\begin{equation}
	\Sigma(k,\omega)=\frac{\gamma^{2}}{L}\sum_{q}\vert\langle u_{0}(k+q)\vert u_{0}(k)\rangle\vert^{2}\overline{G_{00}(k+q,\omega)}
\end{equation}
To the leading order we can ignore the band dispersion induced by
the disorder as we are interested in a flat band limit. Therefore,
we can approximate 
\begin{equation}
	\Sigma(k,\omega)=\frac{\gamma^{2}}{L}\sum_{q}\overline{G_{00}(k+q,\omega)}
\end{equation}
which ignores the effect of the finite dispersion induced by the disorder. 
In this case, we can have $\overline{G_{00}(k,\omega)}\equiv\overline{G_{00}(\omega)}$,
with 
\begin{equation}
	\overline{G_{00}(\omega)}=\frac{1}{\text{\ensuremath{\omega-\Sigma}}(\omega)}\label{eq:greenR}
\end{equation}
which yields a self-consistent equation
\begin{equation}
	\Sigma(\omega)=\frac{\gamma^{2}}{\Omega}\frac{1}{\text{\ensuremath{\omega-\Sigma}}(\omega)},
\end{equation}
where $\Omega$ is the volume of unit cell. For simplicity, we can absorb the coefficients and redefine the $
\gamma^{2}/\Omega\to\gamma^{2}$. Then we can find the solution 
\begin{equation}
	\Sigma(\omega)=\begin{cases}
		\frac{\omega-\sqrt{\omega^{2}-4\gamma^{2}}}{2} & \omega>2\gamma\\
		\frac{\omega-i\sqrt{4\gamma^{2}-\omega^{2}}}{2} & \vert\omega\vert<2\gamma\\
		\frac{\omega+\sqrt{\omega^{2}-4\gamma^{2}}}{2} & \omega<-2\gamma
	\end{cases}
\end{equation}
The disorder has little effect when the $\omega\gg\gamma$ and approximately we have the $\overline{G_{00}(\omega)}=\frac{1}{\omega+i0^{+}}$. For the energy window we are interested in, namely
$\vert\omega\vert\ll2\gamma$, we have 
\begin{align}
	\overline{G_{00}(\omega)} & =\frac{1}{\frac{\omega}{2}+i\gamma}.
\end{align}
Thus, we can extract the relax time $\frac{1}{2\tau}=\gamma$ for
the flat band. 

\subsection{Vertex correction for velocity }

\begin{figure}
	\centering \includegraphics[width=0.5\linewidth]{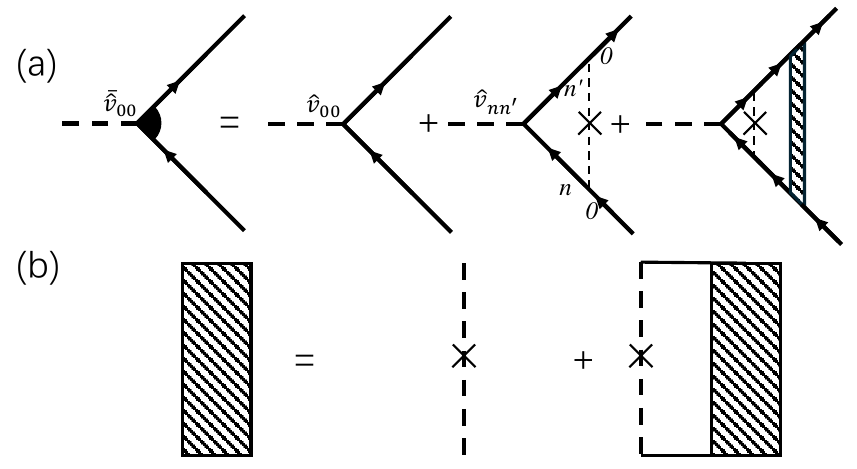} 
	\caption{Feynman diagrams for the vertex correction}
	\label{fig:S-Bethe-Salpeter}
\end{figure}

In the clean limit, the intra-band velocity for the flat band vanishes,
\begin{equation}
	v_{00}(k)=\frac{\partial\epsilon_{0}(k)}{\partial k}c_{0k}^{\dagger}c_{0k}=0.
\end{equation}
One should not expect transport when applying the Kubo-Greenwood formula
to a large system size by ignoring the interface bound states. To
explain the zero-frequency transmission triggered by the disorder,
we then consider the velocity operator renormalized by the disorder.
Before that, from the expression Eq.~(\ref{eq:S-velocityOperator}), 
\begin{align}
	v_{n0}(k) & =\vbra{u_{nk}}\frac{\partial h_{\alpha\beta}(k)}{\partial k}\vket{u_{0k}}\nonumber \\
	& =(\epsilon_{n}(k)-\epsilon_{0}(k))\langle \partial_{k}u_{nk}\vert u_{0k}\rangle,
\end{align}
the interband velocity operator is proportional to the band gap.
Thus, one may expect that its correction may be the leading order
of the order $O(1)$. 

Diagrammatically, in Fig.~\ref{fig:S-Bethe-Salpeter}, we show the relevant Feynman diagram.
In the leading order, we have
\begin{align}
	\overline{v}_{00}(k)= & \int\frac{dq}{2\pi}\sum_{n\neq0}\overline{\Gamma_{0n}(k,q)\Gamma_{00}^{*}(k,q)}G_{nn}(k-q)v_{n0}(k-q)\overline{G_{00}(k-q)}\nonumber \\
	& +\int\frac{dq}{2\pi}\sum_{n\neq0}\overline{\Gamma_{00}(k,q)\Gamma_{n0}^{*}(k,q)}\,\overline{G_{00}(k-q)}v_{0n}(k-q)G_{nn}(k-q)
\end{align}
In details, for the first term we have
\begin{align}
	& \int\frac{dq}{2\pi}\sum_{n\neq0}\overline{\Gamma_{0n}(k,q)\Gamma_{00}^{*}(k,q)}G_{nn}(k-q)v_{n0}(k-q)\overline{G_{00}(k-q)}\nonumber \\
	= & \int\frac{dq}{2\pi}\sum_{n\neq0}\frac{\gamma^{2}}{N\left (\frac{\omega}{2}+i\gamma\right )}\langle u_{0}(k)\vert u_{n}(k-q)\rangle\frac{\epsilon_{n}(k-q)-\cancel{\epsilon_{0}(k-q)}}{\omega-\epsilon_{n}(k-q)}\langle \partial u_{n}(k-q)\vert u_{0}(k-q)\rangle\langle u_{0}(k-q)\vert u_{0}(k)\rangle\nonumber \\
	\to & -\frac{\gamma^{2}}{N\left (\frac{\omega}{2}+i\gamma\right )}\int\frac{dq}{2\pi}\sum_{n\neq0}\langle u_{0}(k)\vert u_{n}(k-q)\rangle\langle \partial u_{n}(k-q)\vert u_{0}(k-q)\rangle\langle u_{0}(k-q)\vert u_{0}(k)\rangle
\end{align}
where the $N$ is the number of unit cells and for the second term we have 
\begin{align*}
	& \int\frac{dq}{2\pi}\sum_{n\neq0}\overline{\Gamma_{00}(k,q)\Gamma_{n0}^{*}(k,q)}\,\overline{G_{00}(k-q)}v_{0n}(k-q)G_{nn}(k-q)\\
	= & \int\frac{dq}{2\pi}\sum_{n\neq0}\frac{\gamma^{2}}{N\left (\frac{\omega}{2}+i\gamma\right )}\langle u_{0}(k)\vert u_{0}(k-q)\rangle\frac{\cancel{\epsilon_{0}(k-q)}-\epsilon_{n}(k-q)}{\omega-\epsilon_{n}(k-q)}\langle \partial u_{0}(k-q)\vert u_{n}(k-q)\rangle\langle u_{n}(k-q)\vert u_{0}(k)\rangle\\
	\to &\frac{\gamma^{2}}{N\left (\frac{\omega}{2}+i\gamma\right )} \int\frac{dq}{2\pi}\sum_{n\neq0}\langle u_{0}(k)\vert u_{0}(k-q)\rangle\langle\partial u_{0}(k-q)\vert u_{n}(k-q)\rangle\langle u_{n}(k-q)\vert u_{0}(k)\rangle\\
\end{align*}
where we use the assumption that the band gap $|\epsilon_0-\epsilon_n|$ is the largest energy scale in comparison to the band width. 
Collect all terms together, we have
\begin{equation}
	\overline{v}_{00}(k)=\frac{2\gamma^{2}}{N\left(\frac{\omega}{2}+i\gamma\right)}\int\frac{dq}{2\pi}\Re \left[\langle u_{0}(k)\vert\partial u_{0}(k+q)\rangle\langle u_{0}(k+q)\vert u_{0}(k) \rangle\right]
\end{equation}
The renormalized velocity is now finite when the disorder effect is
included. Thus, we can expect a finite conductivity. 

\subsection{Diffuson and Ladder approximation}
In the section above, we demonstrated how disorder leads to finite
velocity, which in turn results in finite conductivity. To analyze
diffusion, we will examine the density-density correlator. Unlike
the current-current correlator, the density-density correlator does
not involve inter-band velocity, allowing us to focus on it within
the context of the flat band. We can summarize the vertex correction
using the Bethe-Salpeter equation. For clarity in this section, all
Green functions discussed pertain to the flat band. In this section, we work in a general spatial dimension $d$.

Diffuson describes the behavior of a particle that scatters elastically
off a large number of impurities while traveling through the medium\cite{S-Book-Meso}.
We define the probability of diffusion by taking into account all
possible paths from $\mathbf{r}$ to $\mathbf{r}^{\prime}$ where
the propagating particle scatters elastically off at least one impurity.
Mathematically, we can divide the path from $\mathbf{r}$ to $\mathbf{r}^{\prime}$
into three distinct parts. First, the propagation from the initial
point point until the first scattering event at $\mathbf{r}_{1}$,
then a main part including all scattering events, which is given by
the structure factor $\Gamma_{\omega}(\mathbf{r}_{1},\mathbf{r}_{2})$,
and finally the propagation from the last scattering event at $\mathbf{r}_{2}$
to the endpoint $\mathbf{r}^{\prime}$. Mathematically, we have diffuson $P_{d,\omega}(\mathbf{r},\mathbf{r}^{\prime})$ up to a normalization
factor

\begin{align}
	P_{d,\omega}(\mathbf{r},\mathbf{r}^{\prime}) & =\int d^{d}\mathbf{r}_{1}d^{d}\mathbf{r}_{2}\overline{G_{\epsilon+\omega}^{R}(\mathbf{r},\mathbf{r}_{1})}\overline{G_{\epsilon}^{A}(\mathbf{r}_{1},\mathbf{r})}\Pi({\omega},\mathbf{r}_{1},\mathbf{r}_{2})\overline{G_{\epsilon+\omega}^{R}(\mathbf{r}_{2},\mathbf{r}^{\prime})}\overline{G_{\epsilon}^{A}(\mathbf{r}^{\prime},\mathbf{r}_{2})}\nonumber \\
	& =\overline{G^{R}(\epsilon+\omega)}\overline{G^{A}(\epsilon)}\,\overline{G^{R}(\epsilon+\omega)}\overline{G^{A}(\epsilon)}\Pi({\omega},\mathbf{r},\mathbf{r}')\nonumber \\
	& \equiv P_{0\omega}\Pi(\omega,\mathbf{r},\mathbf{r}^{\prime})P_{0\omega},
\end{align}
with the retarded Green function $\overline{G^{R}(\omega)}=\frac{1}{\frac{\omega}{2}+i\gamma}$
($\omega\ll\gamma$)  and the probability of propagation without any collision $P_{0\omega}=\overline{G^{R}(\epsilon+\omega)}\overline{G^{A}(\epsilon)}$ .
The retarded Green function $\overline{G_{\epsilon+\omega}^{R}(\mathbf{r},\mathbf{r}^{\prime})}$
is given by the Fourier transformation: $\overline{G_{\omega}^{R}(\mathbf{r},\mathbf{r^{\prime}})}=\int\frac{d^{d}\mathbf{k}}{(2\pi)^{d}}\overline{G^{R}(\omega)}e^{i\mathbf{k}\cdot(\mathbf{r}-\mathbf{r^{\prime}})}$.
In the diffusive regime, the time of propagation is much longer than
the scattering time, we have ($\omega\ll\gamma$), 
\begin{align*}
	P_{0\omega} & =\frac{2}{2\gamma^{2}-i\gamma\omega}.
\end{align*}
As for the structure factor $\Pi(\omega,\mathbf{r},\mathbf{r}^{\prime})$,
we can write it recursively as an infinite sum, which is nothing but
the Bethe-Salpeter equation,
\begin{align}
	\Pi(\omega,\mathbf{r},\mathbf{r}^{\prime}) & =\Pi_{0}(\omega,\mathbf{r}-\mathbf{r}^{\prime})+\int d^{d}\mathbf{r}^{\prime\prime\prime}d^{d}\mathbf{r}^{\prime\prime}\Pi_{0}(\omega,\mathbf{r}-\mathbf{r}^{\prime\prime})\overline{G_{\epsilon+\omega}^{R}(\mathbf{r}^{\prime\prime},\mathbf{r}^{\prime\prime\prime})}\overline{G_{\epsilon}^{A}(\mathbf{r}^{\prime\prime\prime},\mathbf{r}^{\prime\prime})}\Pi(\omega,\mathbf{r}^{\prime\prime},\mathbf{r}^{\prime})\nonumber \\
	& =\Pi_{0}(\omega,\mathbf{r}-\mathbf{r}^{\prime})+P_{0\omega}\int d^{d}\mathbf{r}^{\prime\prime}\Pi_{0}(\omega,\mathbf{r}-\mathbf{r}^{\prime\prime})\Pi(\omega,\mathbf{r}^{\prime\prime},\mathbf{r}^{\prime})
\end{align}
with $\Pi_{0}(\omega,\mathbf{r}-\mathbf{r}^{\prime})$ being the bare
vertex, 
\begin{align}
	\Pi_{0}(\omega,\mathbf{r}-\mathbf{r}^{\prime}) & =\int d^{d}\mathbf{q}e^{i\mathbf{q}\cdot(\mathbf{r}-\mathbf{r}^{\prime})}\Pi_{0}(\omega,\mathbf{q})\\
	\Pi_{0}(\omega,\mathbf{q}) & =  \int\frac{d^{d}\mathbf{k}}{(2\pi)^{d}}\overline{\Gamma_{00}(\mathbf{k},\mathbf{q})\Gamma_{00}^{*}(\mathbf{k},\mathbf{q})}\nonumber \\
	&\sim  \gamma^{2}\left(1-\bar{g}_{ij}q_{i}q_{j}\right)\sim\gamma^{2}\frac{1}{1+\bar{g}_{ij}q_{i}q_{j}}
\end{align}
where the $\bar{g}_{ij}=\int \frac{d^{d}\mathbf{k}}{(2\pi)^{d}}g_{ij}(\mathbf{k})$ is the momentum averaged quantum metric and summation over $i,j$ is implicit. For the model used in the main text, we can ignore the local Berry
phase. 
To solve the Bethe--Salpeter equation by ladder approximation, we
can introduce the Fourier transformation 
\begin{equation}
	\Pi(\omega,\mathbf{q})=\int d^d\mathbf{r}e^{-i\mathbf{q}\cdot(\mathbf{r}-\mathbf{r}^{\prime})}\Pi(\omega,\mathbf{r},\mathbf{r}^{\prime})=\int d^d\mathbf{r}e^{-i\mathbf{q}\cdot(\mathbf{r}-\mathbf{r}^{\prime})}\Pi(\omega,\mathbf{r}-\mathbf{r}^{\prime})
\end{equation}
and 
\begin{align}
	& \int d^d\mathbf{r}e^{-i\mathbf{q}\cdot(\mathbf{r}-\mathbf{r}^{\prime})}\int d^{d}\mathbf{r}^{\prime\prime}\Pi_{0}(\omega,\mathbf{r}-\mathbf{r}^{\prime\prime})\Pi(\omega,\mathbf{r}^{\prime\prime},\mathbf{r}^{\prime})\nonumber \\
	= & \int d^d\mathbf{r}e^{-i\mathbf{q}\cdot(\mathbf{r}-\mathbf{r}^{\prime})}\int d^{d}\mathbf{r}^{\prime\prime}\int \frac{d\mathbf{p}d\mathbf{p}'}{(2\pi)^{2d}}e^{i\mathbf{p}\cdot(\mathbf{r}-\mathbf{r}^{\prime\prime})}\Pi_{0}(\omega,\mathbf{p})e^{i\mathbf{p}^{\prime}\cdot(\mathbf{r}^{\prime\prime}-\mathbf{r}^{\prime})}\Pi(\omega,\mathbf{p}^{\prime})\nonumber \\
	= & \int d\mathbf{p}d\mathbf{p}^{\prime}\delta(\mathbf{q}-\mathbf{p})\delta(\mathbf{p}-\mathbf{p}^{\prime})e^{i(\mathbf{q}-\mathbf{p}^{\prime})\cdot\mathbf{r}^{\prime}}\Pi_{0}(\omega,\mathbf{p})\Pi(\omega,\mathbf{p}^{\prime})\nonumber \\
	= & \Pi_{0}(\omega,\mathbf{q})\Pi(\omega,\mathbf{q})
\end{align}
where we assume that the structure factor $\Pi$ is translation-invariant
after the disorder average. Then the Bethe-Salpeter equation is transformed
into 
\begin{equation}
	\Pi(\omega,\mathbf{q})=\Pi_{0}(\omega,\mathbf{q})+P_{0\omega}\Pi_{0}(\omega,\mathbf{q})\Pi(\omega,\mathbf{q})
\end{equation}
We can find the solution easily as
\begin{align}
	\Pi(\omega,\mathbf{q}) & =\frac{\Pi_{0}(\omega,\mathbf{q})}{1-P_{0\omega}\Pi_{0}(\omega,\mathbf{q})}=\frac{1}{\Pi_{0}^{-1}(\omega,\mathbf{q})-P_{0\omega}}\nonumber \\
	& =-\frac{1}{P_{0\omega}-\frac{1}{\gamma^{2}}(1+\bar{g}_{ij}q_{i}q_{j})}.
\end{align}
We are considering the diffusion regime where $\omega\ll\gamma$
and $q\ell_{e}\ll1$.  We can expand $P_{0,\omega}$ to the first order
of $\omega$ 
\begin{align}
	P_{0,\omega}=\frac{2}{2\gamma^{2}-i\gamma\omega}\approx\frac{1}{\gamma^{2}}+i\frac{\omega}{2\gamma^{3}}.
\end{align}
Therefore we obtain the structure factor $\Pi(\omega,\mathbf{q})$
\begin{equation}
	\Pi(\omega,\mathbf{q})=-\frac{1}{i\frac{\omega}{2\gamma^{3}}-\frac{1}{\gamma^{2}}\bar{g}_{ij}(k)q_{i}q_{j}}=-\frac{2\gamma^{3}}{i\omega-2\gamma\bar{g}_{ij}q_{i}q_{j}},
\end{equation}
which diverges at small $\omega$ and $q$. It is easy to find the
diffusion coefficient $D_{ij}$ as 
\begin{equation}
	\label{eq:S-D-and-QM}
	D_{ij}=2\gamma\bar{g}_{ij}.
\end{equation}
In particular, for an isotropic $d$ dimensional case, we have
\begin{align}
	P_{d,\omega}(\mathbf{q}) & =\frac{1}{\mathcal{N}^{2}}\frac{2\gamma}{i\omega-Dq^{2}},\\
	D&= \frac{2\gamma}{d}\mathrm{Tr}\left[\bar{g}_{ij}\right].
\end{align}
The $P_{d,\omega}(\mathbf{r},\mathbf{r}^{\prime},t)$ satisfies the
diffusion equations, 
\begin{equation}
	P_{d}(\mathbf{r},\mathbf{r}^{\prime},t)=\frac{1}{(4\pi Dt)^{d/2}}\exp\left(-\frac{\vert\mathbf{r}-\mathbf{r}^{\prime}\vert^{2}}{4Dt}\right).
\end{equation}
Above, we ignore a normalization factor, and we can recover it by
the normalization condition after shifting $P_{\omega}\rightarrow\frac{1}{\mathcal{N}}P_{\omega}$
\begin{equation}
	P_{\omega}(\mathbf{q}=0)=P_{d,\omega}(\mathbf{q}=0)+P_{0\omega}(\mathbf{q}=0)=\frac{i}{\omega}
\end{equation}
with 
\begin{align}
	P_{\omega}(\mathbf{q}) & =P_{0\omega}P_{0\omega}\Pi(\omega,\mathbf{q})+P_{0\omega}\nonumber \\
	& =P_{0\omega}P_{0\omega}\frac{\Pi_{0}(\omega,\mathbf{q})}{1-P_{0,\omega}\Pi_{0}(\omega,\mathbf{q})}+P_{0\omega}\nonumber \\
	& =\frac{P_{0\omega}}{1-\Pi_{0}(\omega,\mathbf{q})P_{0,\omega}},
\end{align}
and 
\begin{equation}
	P_{\omega}(\mathbf{q})=\frac{1}{P_{0\omega}^{-1}-\Pi_{0}(\omega,\mathbf{0})}=\frac{1}{P_{0\omega}^{-1}-\gamma^{2}}=\frac{1}{\mathcal{N}}\frac{1}{\gamma^{2}-i\gamma\omega/2-\gamma^{2}}=\frac{i}{\omega}
\end{equation}
which gives rise to 
\begin{equation}
	\mathcal{N}=\frac{2}{\gamma}=\frac{1}{\pi\rho_{0}},
\end{equation}
where $\rho_{0}$ is the density of states at $\omega=0$ of the flat
band.


\section{Numerical approaches and results}

\subsection{Transport in two terminal device}
For two terminal case, we can use exact diagonalization to calculate the wavefunction as shown in Fig.~\ref{Fig:S-junction} To effectively perform exact diagonalization, we choose boundary condition
such that: 
\begin{align}
	\psi_{L}(x) & =\psi_{L,0}\sin(kx+\phi_{L}),\\
	\psi_{R}(x) & =\psi_{R,0}\sin[k(x-L-1)].
\end{align}
As such we have $\psi_{R}(L+1)=0$, where $L$ is the length of the
diamond lattice. Assume contact hopping between diamond lattice $S_{L/R}$
on the left/right we have Schr\"odinger equation: 
\begin{equation}
	(H-E)\psi=\begin{pmatrix}-S_{L}\\
		\vdots\\
		0
	\end{pmatrix}.
\end{equation}
Where we have chosen $B_{L}(0)=1$ for convenience. We note that these
equation can be linearly solved via numerical means for arbitrary
energy and give a unique solution due to gauge fixing (except when
$E=0$), and we can obtain the wavefunction for the diamond lattice.
This gives boundary conditions: 
\begin{align}
	\psi_{L,0}\sin\phi_{L} & =1,\\
	t_{N}\psi_{R,0}\sin(k)+S_{R}\psi_{N} & =0,\\
	t_{N}\psi_{L,0}\sin(k+\phi_{L})+S_{L}\psi_{1} & =E.
\end{align}
Where $\psi_{1(N)}$ is the wavefunction of the 1st(Nth) unit cell
in orbital basis. By choosing $S_{L}=S_{R}=S$ we can retrieve information
about the incoming and outgoing wave function from the diamond lattice
wave function: 
\begin{align}
	\psi_{L} & =\frac{2\sqrt{t_{N}^{2}-J\psi_{0}(E-S\psi_{0})}}{4t_{N}^{2}-E^{2}},\\
	\psi_{R} & =-\frac{2J\psi_{N}}{E},\\
	\psi_{L}\sin\phi_{L} & =1.
\end{align}
We can relate this back to transmittance and reflectance by writing:
\begin{align}
	\psi_{L}(x) & =A\sin(kx+\phi_{0})+Ar\sin(kx+\phi_{r}),\\
	\psi_{R}(x) & =At\sin(kx).
\end{align}
Note that due to the gauge choice, the phase of the outgoing wave
is always chosen to be 0. Using energy conservation, namely $t^{2}+r^{2}=1$
we have: 
\begin{equation}
	t=\frac{2\psi_{L}\psi_{R}\sin\phi_{L}}{\sqrt{\psi_{L}^{4}+\psi_{R}^{4}-2\psi_{L}^{2}\psi_{R}^{2}\cos2\phi_{L}}}.
\end{equation}
This correspond to transmittance $\mathcal{T}$: 
\begin{align}
	\mathcal{T} & =|t|^{2}\nonumber \\
	& =\frac{4\psi_{L}^{2}\psi_{R}^{2}\sin^{2}\phi_{L}}{{\psi_{L}^{4}+\psi_{R}^{4}-2\psi_{L}^{2}\psi_{R}^{2}\cos2\phi_{L}}}\nonumber \\
	& =\left[1+\left(\frac{\psi_{L}^{2}-\psi_{R}^{2}}{2\psi_{L}\psi_{R}\sin\phi_{L}}\right)^{2}\right]^{-1}\nonumber \\
	& =\left[1+\left(\frac{\psi_{L}^{2}-\psi_{R}^{2}}{2\psi_{R}}\right)^{2}\right]^{-1}.
\end{align}

\begin{figure}[t]
	\includegraphics[height=0.35\columnwidth]{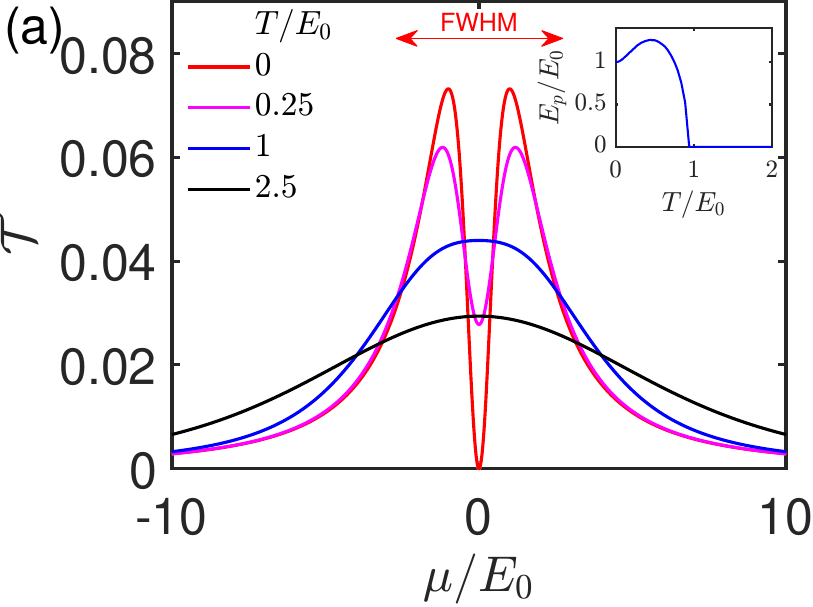}
	\includegraphics[height=0.35\columnwidth]{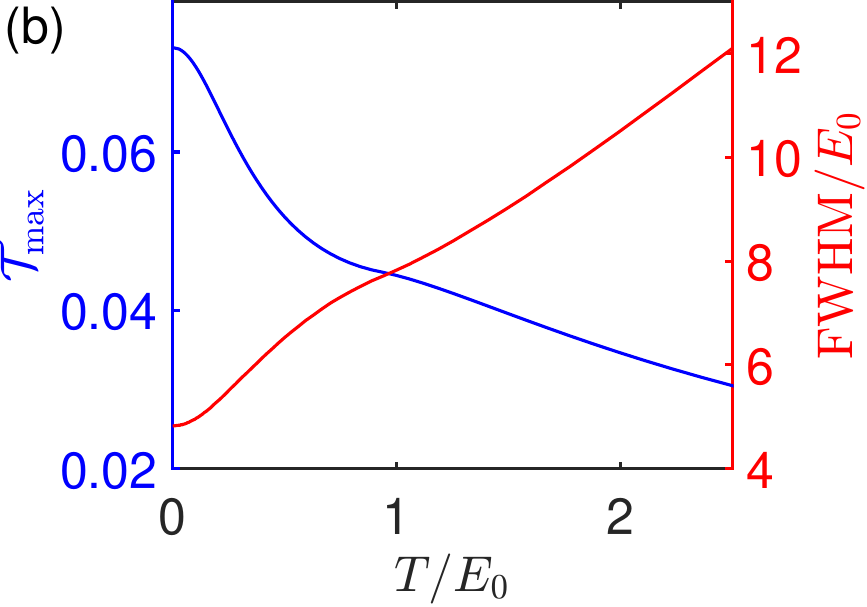} \caption{Finite temperature effect of M/FB/M junction: (a) finite temperature
		transmission profile with the inset on peak location $E_{p}(T)$ and
		(b) FWHM and maximal transmission. At temperature $T=0$, the peaks
		are located at $\pm E_{0}\sim\pm3.7\times10^{-4}$. In (a), there
		remains a drop in the flat-band energy when $T<E_{0}$ and two peaks
		will merge at $T_{0}\sim E_{0}$. In (b), the maximal transmittance
		decreases monotonically as a function of temperature, and the FWHM
		increases instead.}
	\label{fig:S-temperature} 
\end{figure}

We further consider the effect of temperature on transmission. At
temperature $T=\beta^{-1}$, the transmittance $T_{\beta}(\mu_{\mathrm{FB}})$
can be obtained via the Landauer-B\"uttiker formalism $\mathcal{T}_{\beta}(\mu_{\mathrm{FB}})=-\int dE\mathcal{T}(E)\partial_{E}f_{\mathrm{FD}}(E-\mu_{\mathrm{FB}})$
, where $\mu_{\mathrm{FB}}$ is the chemical potential of the diamond
lattice and $f_{\mathrm{FD}}$ is the Fermi-Dirac distribution. First,
finite temperature can lead to improved transmissions for $|\mu_{\mathrm{FB}}|<E_{0}$
where $\mathcal{T}(\mu_{\mathrm{FB}})=16e^{-4L\delta}[(\mu_{\mathrm{FB}}/E_{0})^{2}+\frac{\pi^{2}}{3}(T/E_{0})^{2})]$
for low temperature $T/E_{0}\ll1$. Fig.~\ref{fig:S-temperature}(a)
shows the numerical calculations on the transmission profile, which
agrees with the theoretical predictions. When the temperature $T\sim E_{0}$,
the two peaks will combine into a single one. This process is described
in the inset of Fig.~\ref{fig:S-temperature}(a) about the evolution
of the peak energy $E_{p}$. We also calculate the maximum transmission
$\mathcal{T}_{\mathrm{max}}$ and FWHM in Fig.~\ref{fig:S-temperature}(b)
as we increase in temperature. At low temperature, the FWHM is $\sim4E_{0}$.
As the temperature increases, FWHM gets broadened linearly on temperature.

\begin{figure}[t!]
	\includegraphics[height=0.35\columnwidth]{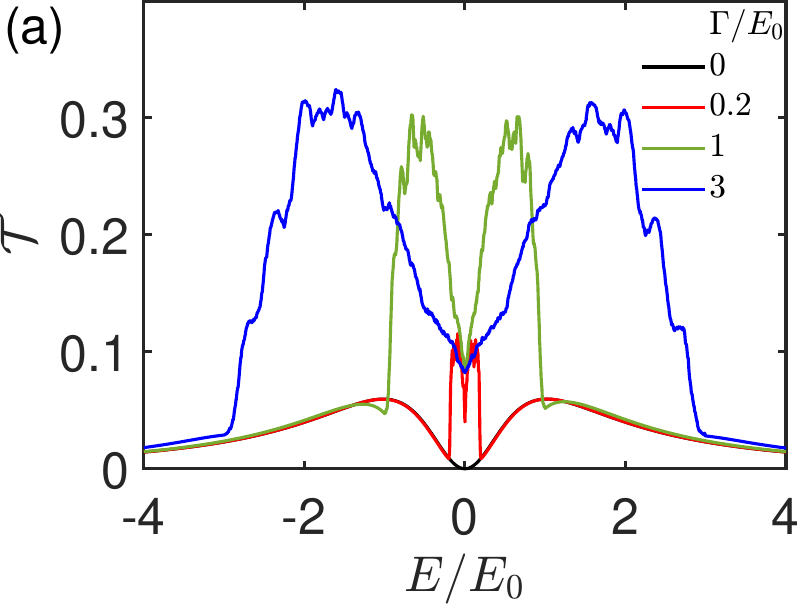} \includegraphics[height=0.35\columnwidth]{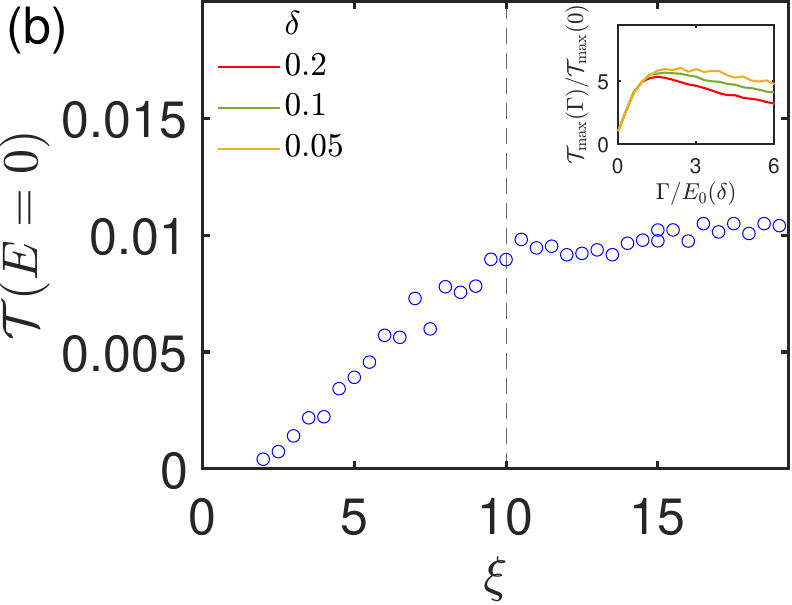}
	\caption{Numerical results on transmission with disorder: (a) Transmission
		profiles at different disorder strengths with parameters $\delta=0.1$
		and $L=10$. The transmittance at energy $|E|<\Gamma$ is enhanced,
		while the disorder has almost no effect for $|E|>\Gamma$. (b) The
		transmittance $\mathcal{T}$ at $E=0$. We increase the localization
		length $\xi$ by tuning $\delta$ while fixing by fixing disorder
		strength $\Gamma=0.05t_{N}$ and $L=20$. The transmittance increases
		linearly with the quantum metric length when $L\delta<1$, with the
		dash vertical line in (b) marking the position $L\delta=1$. Beyond
		$L\delta=1$, $\mathcal{T}(E=0)$ remains roughly constant. In the
		inset of (b), the maximal transmittance $\mathcal{T}_{\mathrm{max}}(\Gamma)$
		of the transmission profiles as a function of disorder strength $\Gamma$
		for $\delta=0.05,0.1$ and $0.2$ with $L\delta=1$. At weak disorder
		$\Gamma/E_{0}(\delta)<1$ with $E_{0}(\delta)=4T_{\partial}^{2}\delta/t_{N}$,
		the maximal transmittance shows a unified linear dependency with normalized
		disorder strength $\Gamma/E_{0}$. At strong disorder $\Gamma/E_{0}\gg1$,
		the transmission is suppressed as $\sim1/\Gamma$ when we further
		increase disorder strength. }
	\label{fig:S-disorder} 
\end{figure}


We can proceed to establish an understanding on the effects of the
disorder. The bound states, which in the clean limit only exist and
is localized at the two interfaces, can be excited by disorders to
emerge and propagate within the diamond lattice. Numerically, we can
examine the transmission when introducing the disorder terms $H_{\text{dis}}=\sum_{i}\sum_{\alpha=abc}w_{i}\alpha_{i}^{\dagger}\alpha_{i}$
on the diamond lattice in the exact diagonalization at zero temperature,
where $w_{i}$ is the random chemical potential $w_{i}\in[-\Gamma,\Gamma]$
of a uniform distribution.

Essentially, disorder can break the degeneracy of the flat band, which
gives rise to a distribution of energy levels $E\in[-\Gamma,\Gamma]$,
and a bandwidth $W\sim2\Gamma$. In Fig.~\ref{fig:S-disorder}(a), we have illustrated
the transmission profile for different disorder strengths. We find
that the transmittance increases for $\abs{E}<\Gamma$ while for $\abs{E}\gg\Gamma$,
the effect of disorder is insignificant, and the transmission profile
is similar to the clean limit. Contrary to the disorder-free case,
the transmittance $\mathcal{T}(E=0)$ at zero energy becomes a finite
value. Interestingly, as shown in Fig.~\ref{fig:S-disorder}(b), the
transmittance $\mathcal{T}(E=0)$ first increases linearly and then
approaches a constant value when we increase the localization length
of the bound states by fixing $L=20$ and the disorder strength $\Gamma\gg E_{0}$.
This differs from the conventional case of one-dimensional single
dispersive band, where transport is suppressed in the dirty limit,
due to reduction in the meanfree path by disorder.

Another aspect of the enhancement can be inferred from the maximal
transmittance $\mathcal{T}_{\text{max}}(\Gamma)$, which is depicted
in the inset of Fig. \ref{fig:S-temperature}(b) of which we keep $L\delta=1$ as suggested
by Eq. \eqref{im_pair}. When $\Gamma<\Gamma_{0}$ where $\Gamma_{0}$ is the
optimal disorder strength for the peak of maximal transmittance, we
have universal behavior $\mathcal{T}_{\text{max}}(\Gamma)/\mathcal{T}_{\text{max}}(0)=1+\alpha\Gamma/E_{0}(\delta)$
with $E_{0}(\delta)=4T_{\partial}^{2}\delta/t_{N}$ and $\alpha\approx5$
is independent of $\delta$. This justifies that $E_{0}$ works as
the natural energy scale for the flat-band junction. In particular,
at the disorder strength $\Gamma_{0}$, we observed an enhancement
of the maximal transmittance of up to 5 times, in comparison with
the clean limit.

\subsection{Transport in four terminal device}


To explore the disorder induced delocalization in the bulk, we are
considering the four terminal device, suppose the coupling Hamiltonian
from lead $\alpha$ to central is $\tau_{C,\alpha}$ with the central
Hamiltonian $H_{CC}$ unchanged, we can write down the total Hamiltonian
as \cite{S-Multiterminal-Cal}
\begin{equation}
	\mathcal{H}=\begin{pmatrix}H_{1} & 0 & \cdots & 0 & 0 & \tau_{C,1}^{\dagger}\\
		0 & H_{2} & \cdots & 0 & 0 & \tau_{C,2}^{\dagger}\\
		\vdots & \vdots & \ddots & \vdots & \vdots & \vdots\\
		0 & 0 & \cdots & H_{N-1} & 0 & \tau_{C,N-1}^{\dagger}\\
		0 & 0 & \cdots & 0 & H_{N} & \tau_{C,N}^{\dagger}\\
		\tau_{C,1} & \tau_{C,2} & \cdots & \tau_{C,N-1} & \tau_{C,N} & H_{CC}
	\end{pmatrix}.
\end{equation}
Notice that every block Hamiltonian of semi-infinite lead $H_{\alpha}$
are infinite dimensional matrices. Follow the procedure above, we
first solve the Green's function of the central Hamiltonian $(E-H_{CC}+i\eta)^{-1}$,
then we need to calculte the self energy correction. Define the block
matrix 
\begin{equation}
	\tau=(\tau_{C,1},\tau_{C,2},\cdots,\tau_{C,N-1},\tau_{C,N}),\quad E-H_{\text{Lead}}=\begin{pmatrix}E-H_{1} & 0 & \cdots & 0 & 0\\
		0 & E-H_{2} & \cdots & 0 & 0\\
		\vdots & \vdots & \ddots & \vdots & \vdots\\
		0 & 0 & \cdots & E-H_{N-1} & 0\\
		0 & 0 & \cdots & 0 & E-H_{N}
	\end{pmatrix},
\end{equation}

so that we can write the total Hamiltonian into a simpler form 
\begin{equation}
	E-\mathcal{H}=\begin{pmatrix}E-H_{\text{Lead}} & \tau^{\dagger}\\
		\tau & E-H_{CC}
	\end{pmatrix}.
\end{equation}
Apply the inverse to the $2\times2$ matrix and we can get 
\begin{equation}
	\mathcal{G}(E)=(E-\mathcal{H})^{-1}=\begin{pmatrix}G_{\text{Lead}}(E)+G_{\text{Lead}}(E)\tau^{\dagger}\mathbf{G}^{R}\tau G_{\text{Lead}}(E) & -G_{\text{Lead}}(E)\tau^{\dagger}\mathbf{G}^{R}\\
		-\mathbf{G}^{R}\tau G_{\text{Lead}}(E) & \mathbf{G}^{R}
	\end{pmatrix}
\end{equation}
where $G_{\text{Lead}}(E)=(E-H_{\text{Lead}})^{-1}$, and $\mathbf{G}^{R}(E)=(E-H_{CC}-\tau G_{\text{Lead}}\tau^{\dagger})^{-1}$.
Notice that the $(E-H_{\text{Lead}})$ is block diagonal, so we can
write the Green's function as 
\begin{equation}
	G_{\text{Lead}}(E)=(E-H_{\text{Lead}})^{-1}=\begin{pmatrix}g_{1} & 0 & \cdots & 0 & 0\\
		0 & g_{2} & \cdots & 0 & 0\\
		\vdots & \vdots & \ddots & \vdots & \vdots\\
		0 & 0 & \cdots & g_{N-1} & 0\\
		0 & 0 & \cdots & 0 & g_{N}
	\end{pmatrix}.
\end{equation}
Use the multiplication of block matrix 
\begin{equation}
	\tau G_{\text{Lead}}\tau^{\dagger}=\sum_{n=1}^{N}\tau_{C,n}g_{n}\tau_{C,n}^{\dagger}=\sum_{n=1}^{N}\Sigma_{n},\quad\Sigma_{n}=\tau_{C,n}g_{n}\tau_{C,n}^{\dagger}.
\end{equation}
So we can write the final Green's function with self energy correction
as 
\begin{equation}
	G_{CC}(E)=\left(E-H_{CC}-\sum_{n=1}^{N}\Sigma_{n}(E)\right)^{-1}
\end{equation}
Now, we can use the Fisher-Lee relation with the corresponding spread
$\Gamma_{\alpha}$ corresponds to required terminal $\alpha$ and
get the transmission 
\begin{equation}
	\mathcal{T}_{\alpha\beta}=\trace[\Gamma_{\alpha}\mathbf{G}^{R}\Gamma_{\beta}(\mathbf{G}^{R})^{\dagger}]
\end{equation}

The M/FB/M junction consists of a diamond lattice connected to metallic leads. The total Hamiltonian is
\begin{align}
	\hat{H}=\hat{H}_{\text{diamond}}+\sum_{I}\hat{H}_{\text{M}_{I}}+\hat{H}_{c}.
\end{align}
Here, $\hat{H}_{\text{diamond}}$ describes the diamond lattice. As shown in the main text Fig. 1(a), it contains three orbitals (A, B, and C) per unit cell, with annihilation operators $\hat{a}_{x}$, $\hat{b}_{x}$, and $\hat{c}_{x}$. The Hamiltonian is
\begin{align}
	\hat{H}_{\text{diamond}}&=\sum_{x}J_{+}(\hat{b}_{x}^{\dagger}\hat{a}_{x}^ {}+\hat{c}_{x}^{\dagger}\hat{a}_{x})+J_{-}(\hat{a}_{x}^{\dagger}\hat{b}_{x+1}+\hat{c}_{x}^{\dagger}\hat{a}_{x+1})\nonumber\\
	&+\text{H.c.}-\mu_F\sum_{\alpha=abc}\hat{\alpha}^\dagger_i\hat{\alpha}_i
\end{align}
where $J_{\pm} = J(1 \pm \delta)$ denotes hopping strength, $x$ is the unit cell index, and $\mu_F$ is the chemical potential. The diamond lattice features one flat band and two dispersive bands, with the flat band separated by a gap $\Delta = 2\sqrt{2}J\delta$. The quantum metric for Bloch state $\vket{u(k)}$ is defined as
\begin{align}
	\mathcal{G}(k) & = \langle\partial_{k}u(k)|(1-|u(k)\rangle\langle u(k)|)|\partial_{k}u(k)\rangle,
\end{align}
and the averaged quantum metric over the Brillouin zone of the flat-band Bloch state $|u_0(k)\rangle$ is
\begin{equation}\label{eq:S-qm}
	\overline{\mathcal{G}}=\frac{1}{2\pi}\int_{-\pi}^\pi\mathcal{G}(k)dk=\frac{a(1-\delta)^2}{8\delta},
\end{equation}
where $a$ is the lattice constant. 
We set $a=1$ throughout. 

The second term $\hat{H}_{\text{M}_I}$ describes the Hamiltonian of semi-infinite leads $\text{M}_I$ ($I=1,2,3,4$). To minimize finite-size effects, the lattice of leads 1 and 4 matches the central diamond lattice without disorder, while leads 2 and 3 are modeled as metallic wires with nearest-neighbor hopping $t_N$ and chemical potential $\mu_N$:
\begin{equation}
	\hat{H}_{\text{M}_I}=\sum_{\langle ij \rangle\in \text{M}_I}t_{N}(\hat{\gamma}_{i}^{\dagger}\hat{\gamma}_{j}+\text{H.c.})-\mu_{N}\sum_{i\in \text{M}_I}\hat{\gamma}_{i}^{\dagger}\hat{\gamma}_{i},    
\end{equation}
where $\langle ij \rangle$ denotes nearest-neighbor hopping. We set $\mu_N = \mu_F = 0$ to align the Fermi energy of the leads with the flat band. The third term, $\hat{H}_c$, describes the coupling between the diamond lattice and the metallic leads $\text{M}_2,\text{M}_3$ with strength $T_{\partial}$:
\begin{equation}
	H_c = T_{\partial} \sum_{i\in\{\partial\text{M}_2,\partial\text{M}_3\}}\sum_\alpha (\hat{\gamma}_{i}^\dagger \hat{\alpha}_i + \text{H.c.}),
\end{equation}
where the $\alpha$ labels orbital A, B, and C at the coupling position. subscripts indicate the connecting terminals as shown in Fig.~\ref{Fig:S-junction}. For four-terminal measurements, we employ three-channel metallic leads connected to each orbital at the central disordered diamond lattice. 

In all numerical calculations, we fix $J\delta=10$, yielding a gap $\Delta=20\sqrt{2}$ and keep the disorder strength $\Gamma\ll \Delta$ to preserve the flat band isolated from other dispersive bands. The coupling is set to $T_{\partial}=0.1$ to simulate the imperfect connection and $t_{N}=1$ serves as the energy unit throughout.


\subsection{Wave packet dynamics for 1D diamond lattice}
\begin{figure}[H]
	\centering \includegraphics[width=0.48\columnwidth]{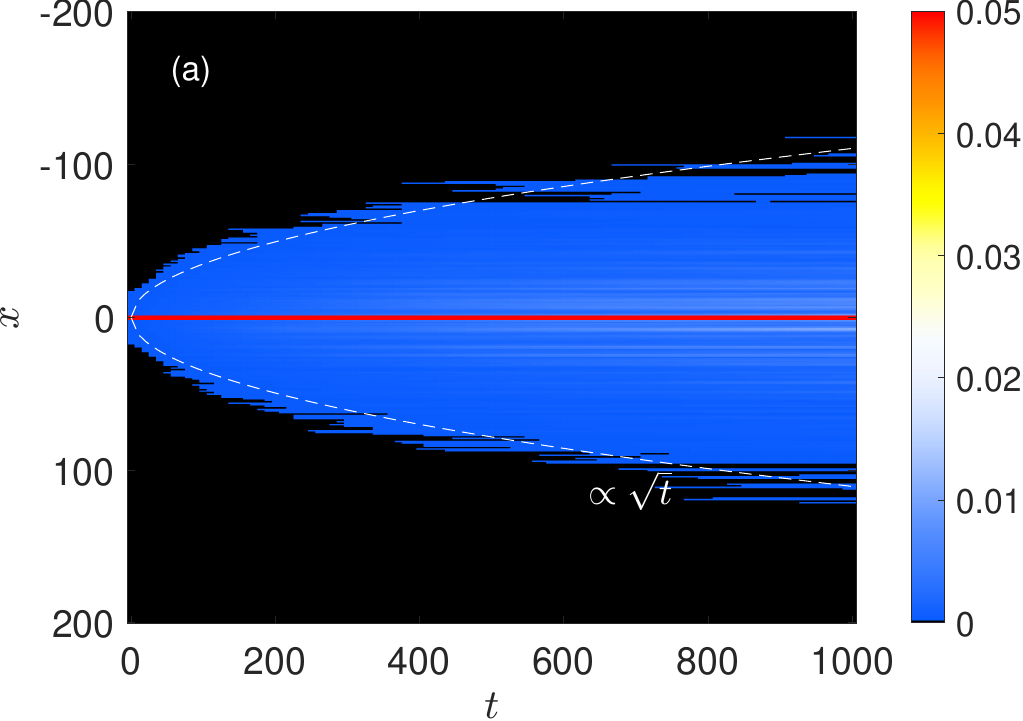} 
	\includegraphics[width=0.48\columnwidth]{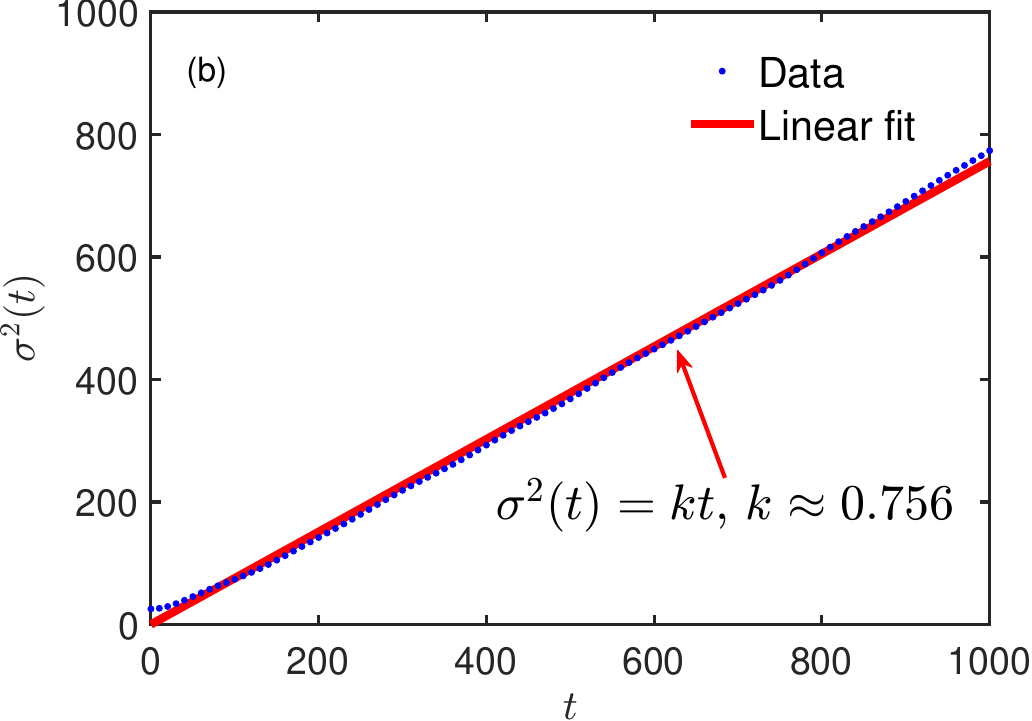} 
	\caption{(a) Time evolution of the unit cell occupation $n_i(t)$ for the wave packet $\vket{\psi(t)}$. The simulation uses a chain of length $L=401$ with parameters $\Gamma=0.1, \delta=0.01$ averaged over 30 disorder realizations. The evolution time is restricted to $t\le100/\Gamma$. (b) Mean square distance $\Delta X^2(t)$ calculated using Eq.(\ref{eq:S-MSD}). The slope of the linear fit, $k=0.75625\pm0.0032$, corresponds to a diffusion coefficient of  $D=k/2\approx0.3781$.}
	\label{Fig:S-Diffusion} 
\end{figure}

To illustrate the relationship between the diffusion coefficient and the quantum metric as described in Eq. (\ref{eq:S-D-and-QM}), we utilize the mean square displacement $\Delta X^2(t)$ derived from the time-evolved wavefunction \cite{S-Diffusion-MSD-1}. Consider a pure one-dimensional diamond chain of length $L$ with open boundary conditions. The Hamiltonian of the system is given by: \begin{equation}
	H=H_0+V,
\end{equation}
where $H_0$ represents the Hamiltonian of the diamond lattice, and $V$ denotes the onsite disorder potential. The tight-binding Hamiltonian $H_0$ can be numerically diagonalized to obtain the eigenstates $|\psi_i\rangle$. From these, we select the flat-band states and construct the projector:
\begin{equation}
	P_F=\sum_F |\psi_F\rangle\langle\psi_F|,
\end{equation}
which allows us to exclude contributions from dispersive states.
Let $|\phi_{i\alpha}\rangle $ denote the wavefunction of state $\vket{\phi}$ at site $\alpha$ in the $i$-th unitcell. Setting the central unit cell as the origin, we prepare the initial wavefunction $\vket{\phi}$ such that $|\phi_{0B}(t=0)\rangle =|\phi_{0C}(t=0)\rangle= 1/\sqrt{2}$ in the central unit cell of the diamond lattice. We then project out the dispersive states to obtain the wave packet $\vket{\psi}$:
\begin{equation}
	\vket{\psi} = \frac{\vket{\psi_0}}{\sqrt{\vbra{\psi_0}\ket{\psi_0}}},\quad\vket{\psi_0}=P_F\vket{\phi}
\end{equation}
Next, we evolve the state according to the full Hamiltonian via $\vket{\psi(t)}=e^{-iHt}\vket{\psi}$ and define the mean square displacement as:
\begin{equation}
	\label{eq:S-MSD}
	\Delta X^2(t) = \average{x^2(t)}-\average{x(t)}^2=\sum_{i=-L/2}^{L/2}i^2 n_i(t)-\left(\sum_{i=-L/2}^{L/2}i\, n_i(t)\right)^2,
\end{equation}
where $n_i(t)=\sum_{\alpha=ABC}\vbra{\psi_{i\alpha}(t)}\ket{\psi_{i\alpha}(t)}$. 
For diffusive transport, it can be shown that the mean square displacement follows \cite{S-Diffusion-MSD-1,S-Diffusion-MSD-2}
\begin{equation}
	\label{eq:S-diffusion_MSD}
	\Delta X^2(t)=2Dt,
\end{equation}
where $D$ is the diffusion coefficient. This relationship allows us to extract $D$ by fitting $\Delta X^2(t)$ to a linear function. The results are shown in Fig.~\ref{Fig:S-Diffusion}(b), where the mean square displacement exhibits a linear growth with time. By determining the slope of this linear behavior, we obtain the diffusion coefficient $D$.

As demonstrated in Eq. (\ref{eq:S-D-and-QM}), the diffusion coefficient $D$ is a function of disorder strength $\gamma$ and the quantum metric length $\bar{g}=1/8\delta$. Due to a discrepancy between the numerical disorder strength $\Gamma$ and the theoretical $\gamma$ (up to a constant factor), the diffusion coefficient satisfies:
\begin{equation}
	\label{eq:S-diffusion}
	D=2\gamma \bar{g}=C\times \Gamma \bar{g}.
\end{equation}
where $C$ is a proportionality constant that can be determined self-consistently. To find $C$, we fix the disorder strength at $\bar{\Gamma}$ and calculate the diffusion coefficient $D(\delta,\bar{\Gamma})$ for varying $\delta$. This yields $D(\delta,\bar{\Gamma})=k\times\bar{g}$, where the constant $C$ is given by $C=k/\bar{\Gamma}$. For $\bar{\Gamma}=0.1$ we find $C=0.337$. This coefficient is then used to validate Eq.(\ref{eq:S-D-and-QM}) across different parameters, as summarized in Table \ref{tab:S-tab1}.
\begin{table}[H]
	\caption{Diffusion coefficients obtained from theoretical predictions and numerical simulations.}
	\label{tab:S-tab1}
	\begin{center}
		\begin{tabular}{ccccc}
			\hline
			$J$& $\delta$ & $\Gamma$ & $D_\text{pred}$ & $D_\text{numeric}$ \\ 
			\hline
			1000 & 0.1 & 0.1 & $0.0421$ & $0.0182$\\ 
			\hline
			1000 & 0.1 & 0.01 & $0.0042$ & $0.0026$\\ 
			\hline
			1000 & 0.01 & 0.1  & $0.4213$ & $0.4338$ \\ 
			\hline
			1000 & 0.05 & 0.07  & $0.0590$ & $0.0442$ \\ 
			\hline
			10000 & 0.01 & 0.1  & $0.4213$ & $0.3744$ \\ 
			\hline
			100000 & 0.01 & 0.1  & $0.4213$ & $0.4184$ \\ 
			\hline
			100000 & 0.03 & 0.2  & $0.2808$ & $0.2493$ \\ 
			\hline
		\end{tabular}
	\end{center}
\end{table}

Table.~\ref{tab:S-tab1} summarizes the diffusion coefficients obtained from Eq.(\ref{eq:S-diffusion}) and numerical fitting of eq.(\ref{eq:S-diffusion_MSD}). The system parameters are listed, and all data are computed for a chain of length $L=1001$, averaged over 20 disorder realizations.

\begin{figure}[H]
	\centering \includegraphics[width=0.48\columnwidth]{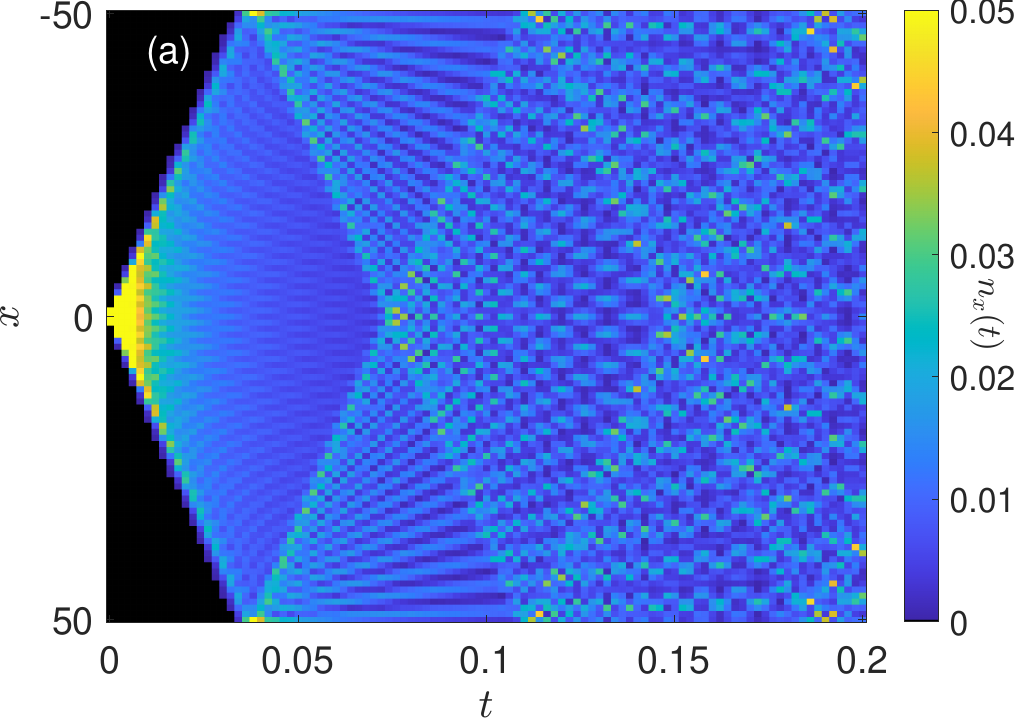} 
	\includegraphics[width=0.48\columnwidth]{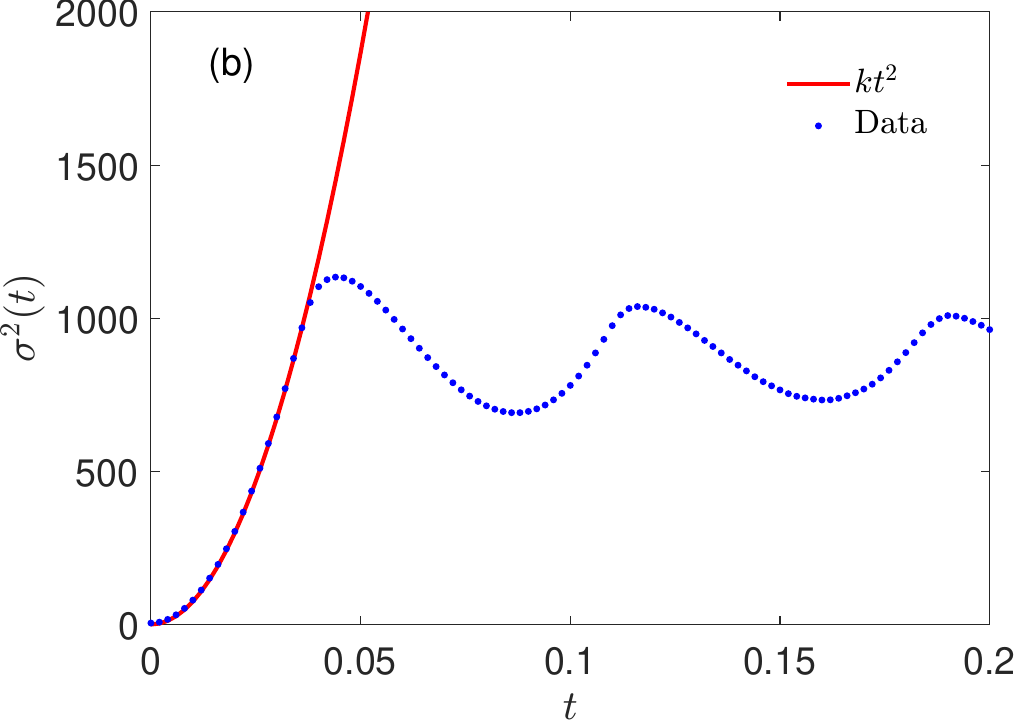} 
	\caption{(a) Time evolution of the unit cell occupation $n_i(t)$ for the wave packet $\vket{\psi'(t)}$. The simulation uses a chain of length $L=101$ with parameters $\delta=0.01$ without disorder. (b) Mean square distance $\Delta X^2(t)$ calculated using Eq.(\ref{eq:S-MSD}). The coefficient for the quadratic fit $k=(7.468\pm0.056)\times 10^5$}
	\label{Fig:S-Diffusion-Dispersive} 
\end{figure}

For comparison, we also simulate the time evolution of the wave packet $\vket{\psi'(t)}=e^{-iHt}\vket{\psi'_0}$, where $\vket{\psi_0'}$ is composed of dispersive states in Fig.~\ref{Fig:S-Diffusion-Dispersive} without disorder:
\begin{equation}
	\vket{\psi'}=\frac{\vket{\psi_0'}}{\sqrt{\vbra{\psi_0'}\ket{\psi_0'}}},\quad\vket{\psi_0'}=(\mathbbm{1}-P_F)\vket{\phi}
\end{equation}
The ballistic motion can be seen clearly at the beginning of the evolution and the MSD has a quadratic profile.

%


\begin{thebibliography}{72}
\bibitem [{QM-()}]{QM-Addedcontent}%
	\BibitemOpen
	\href@noop {} {}\bibinfo {note} {When we tune the quamtum metric length to be
		larger, the DC transport will drop in turn. This may due to the Wannier wave
		function overlap are comparable to the junction length, while the disorder
		was distributed randomly around unit cell. Since the disorder strength is
		fixed, the large overlap will the cover many unit cells and the disorder was
		averaged to zero inside the overlap, results in suppressed DC
		transport.}\BibitemShut {Stop}%
\end{thebibliography}
\end{document}